\documentclass[aps,pra,twocolumn,showpacs,floatfix]{revtex4-1}
\usepackage{amssymb}
\usepackage{amsmath}
\usepackage{graphicx}
\usepackage{dcolumn}
\usepackage{bm}
\usepackage[]{hyperref}
\usepackage{color}

\newcommand{\GammaF}{\Gamma_{\hspace{-1.5pt}F}}

\newcommand{\bloch}{\psi_{\scalebox{0.6}{$B$}}}
\newcommand{\ket}{\rangle}
\newcommand{\bra}{\langle}
\newcommand{\BZ}{\scalebox{0.6}{BZ}}
\newcommand{\spc}{|}
\newcommand{\sinc}{\mbox{sinc}}
\newcommand{\C}[1]{{\mathcal{#1}}}
\newcommand{\R}{\mathrm}
\newcommand{\s}[1]{{\hspace{#1pt}}}

\newsavebox{\Wbox}
\savebox{\Wbox}{\scalebox{0.6}{W}}

\definecolor{correction}{RGB}{150,120,100}

\newcommand{\BOnew}{DEBO}

\begin{document}

\title{Delocalization-enhanced Bloch oscillations and driven resonant tunneling in optical lattices for precision force measurements}

\author{M.~G.~Tarallo} 
\author{A.~Alberti}
\altaffiliation{Institut f\"ur Angewandte Physik, Universit\"at Bonn, Wegelerstr.\ 8, 53115 Bonn, Germany}
\author{N.~Poli}
\author{M.~L.~Chiofalo} 
\altaffiliation{Department of Mathematics and INFN, University of Pisa, Largo B.~Pontecorvo 5, 56127 Pisa, Italy}
\author{F.-Y.~Wang}
\author{G. M.~Tino} 
\email{Guglielmo.Tino@fi.infn.it}

\affiliation{Dipartimento di Fisica e Astronomia and LENS -- Universit\`a di Firenze,\\
INFN -- Sezione di Firenze, Via Sansone 1, 50019 Sesto Fiorentino, Italy}

\begin{abstract}
In this paper we describe and compare different methods used for accurate determination of forces acting on matter-wave packets in optical lattices. The quantum interference nature responsible for the production of both Bloch oscillations and coherent delocalization is investigated in detail. We study conditions for optimal detection of Bloch oscillation for a thermal ensemble of cold atoms with a large velocity spread. We report on the experimental observation of resonant tunneling in an amplitude-modulated (AM) optical lattice up to the sixth harmonic with Fourier-limited linewidth. We then explore the fundamental and technical phenomena which limit both the sensitivity and the final accuracy of the atomic force sensor at $10^{-7}$ precision level~\cite{Poli2011}, with an analysis of the coherence time of the system and addressing few simple setup changes to go beyond the current accuracy.

\end{abstract}


\maketitle

\section{Introduction}
\label{sec:Intro}

Coherent manipulation of cold atoms trapped in optical lattice potentials finds powerful applications in different fields like high resolution metrology~\cite{Ye27062008}, quantum information processes~\cite{Bloch:2008p1810}, strongly correlated quantum phases~\cite{Greiner:2002p709,Kinoshita:2004p774,Bloch:20081173}, as much as in the determination of forces at micrometer resolution~\cite{Anderson:1998p517,Roati:2004p1040,Clade:2005p3137, Carusotto:2005p587,Ferrari06,Wolf:2007p986}. In particular, loading cold atoms into tilted optical lattices led to the possibility to detect Bloch oscillations in the presence of constant forces~\cite{Bloch29,BenDahan:1996p31}, as well as Wannier-Stark ladders and tunneling~\cite{Raizen:1997p753}, while the introduction of dynamical driving of the lattice depth has been used for several purposes: performing spectroscopy studies of interband excitations~\cite{Friebel:1998p575,Denschlag:2002p757,PhysRevLett.107.135303}, characterizing  the Mott-insulator regime~\cite{Stoferle:2004p801}, quantum simulation{s of} spin chains~\cite{PhysRevLett.107.210405,Simon2011} and demonstrat{ing} Loschmidt echoes in thermal gases~\cite{Cucchietti:10,Alberti:2010p70} and their utilization as a high fidelity atom mirror~\cite{Alberti:2010p70}. 

In the present paper we aim to show how it is possible to merge the two phenomena of Bloch oscillations in tilted optical lattices and dynamical driving of lattice phase and amplitude to perform measurements  of forces with high sensitivity and accuracy. This combination leads to the realization of novel technique that we call delocalization-enhanced Bloch oscillations (\BOnew), which is characterized in detail both theoretically and experimentally in this work. We provide for the first time an unitary treatment of the dynamical system originated by an accelerated periodic potential which is driven by both amplitude (AM) or phase (PM) modulation. We give comprehensive theoretical explanation of the effect originating from an interplay between external acceleration and the driving fields. This theoretical analysis and its predictions can be used to understand and interpret the experimental demonstrations about controlling and tailoring quantum transport of matter waves through coherent delocalization by resonant phase~\cite{Ivanov:2008p37} or amplitude~\cite{Alberti:2010p70} modulation, observation of quantum transport over macroscopic distances, macroscopic Bloch oscillations in real space by off-resonant modulation~\cite{Alberti:2009p45}, and coherent phase imprint over the matter waves showing Loschmidt echoes. Furthermore, starting from the analysis of space and time symmetries, we provide a demonstration of the resonance spectrum accuracy of the coherent delocalization technique.

Particular attention is then paid to the utilization of these techniques to develop a force sensor for precision measurements of the gravitational acceleration $g$ based on an accurate determination of Bloch frequency. We provide experimental measurements of Bloch frequency by direct observation of the modulation frequency at which resonant tunneling occurs, and we compare it with the one derived from \BOnew~technique in momentum space after several thousands periods. Since these two techniques rely on coherences of matter waves, we experimentally investigate the decoherence phenomena limiting the force measurement sensitivity. Finally we review all the important systematic effects and the specific impact on the two different techniques, providing the error budget and thus, for the first time, a significant comparison of their performance.

The paper is organized as follows. We outline in Sec.~\ref{sec:Theory} the important aspects characterizing the quantum motion in modulated optical lattices in terms of Wannier-Stark states, discussing it from the two alternative points of view of position and momentum space. We calculate the transport resonance spectrum by means of the Floquet theory~ \cite{Grifoni:1998tk}. We provide an analytical simple model explaining the interference peak enhancement of Bloch oscillation visibility resulting from prior lattice modulation. We describe in Sec.~\ref{sec:ExperimentalSetup} the experimental setup and procedures adopted in order to tune the governing parameters and study the system's observables. We explicitly discuss in Sec.~\ref{sec:SensitivityGeneral} the general advantages of using modulation driving to improve phase sensitivity in the measurement of Bloch oscillations frequency. The analysis and  comparison of sensitivity obtained with Bloch oscillation and resonant tunneling measurements is presented in Sec.~\ref{sec:PS_BO}, where the error budget is analyzed. Sec.~\ref{sec:Conclusions} finally contains concluding remarks and perspectives on how to further improve sensitivity for precision force measurements.

\section{Theoretical framework: quantum motion in modulated optical lattices}
\label{sec:Theory}

\subsection{The system: tilted optical lattices}

We consider the quantum motion {of} non-interacting {ultracold atoms of} mass $M$, loaded into the one-dimensional (1D) optical lattice $U(z)$
\begin{equation}\label{eq:Uz}
U(z) = - \frac{U_0}{2}\cos\left(\frac{2\pi}{d}z\right )
\end{equation}
with lattice constant $d$ and lattice depth $U_0$  \cite{Morsch:2006em}. This is originated by the interference pattern of two counter-propagating laser beams with wavelength $\lambda_L=2\pi/k_L$;  {the interference contrast controls directly the depth $U_0$, while the wavelength determines the lattice period $d=\lambda_L/2$}. As we are interested in {a} non-dissipative potential, the laser beam frequency {has} to be far detuned from any atomic transition {to make} the scattering rate at which atoms spontaneously emit photons {negligible}.
 
{Motional} degrees of freedom can be driven by external forces {$F(z) = -\partial_z V(z)$} and quantum motion can be tailored {by} modulating the optical lattice either in amplitude or in phase. The effective Hamiltonian governing the system is
\begin{equation}\label{eq:startH}
\C{H}(z,p,t) =  \frac{p^2}{2 M}+U(z)[1+\alpha f(t)] - \beta z
f(t)+V(z)\,,
\end{equation}
where $f(t)=\sin(\omega_Mt-\phi)$ is the time-dependent harmonic {driving} function, $\alpha$ and $\beta$ are respectively the amplitude and phase modulation depth{s.} In the reference frame co-moving with the lattice (Kramers-Henneberger transformation), $\beta$ is a function of the spatial driving amplitude $z_0$ and $\omega_M$ as $\beta=M\hspace{1pt}z_0\hspace{1pt}\omega_M^2$. 

In the absence of both external forces and modulations, the stationary solutions of (\ref{eq:startH}) are Bloch states $\spc \bloch^{j}(k)\ket$ labelled by band index $j$ and quasi-momentum $k$, which satisfy Bloch's theorem expressing invariance after translation of $n$ lattice sites \cite{Bloch29}. The corresponding energy spectrum consists of Bloch bands $E_{j}(k)$. These are conveniently represented within the first Brillouin zone $\mathrm{BZ}=[-k_L,k_L]$, so that each band of index $j$ is continuous over this interval. Within a semiclassical approach, the derivative of the energy band determines as a function of $k$ the group velocity of the wave packets in the optical lattice:
\begin{equation}\label{eq:vg}
v_g (k)=\frac{1}{\hbar}\frac{\partial E_{j}(k)}{\partial k}\,.
\end{equation}

Depending on the particular quasi-momentum $k$ distribution, an initially localized  particle can propagate from site to site by means of the so-called Bloch tunneling, resulting in a ballistic spread over the whole lattice \cite{Visser:1997p163}. By applying external forces and lattice modulations, it is possible to control and coherently drive the quantum motion of wave packets.

In the simplest case of a constant force $F(z)=F$, atoms undergo oscillations with angular frequency
\begin{equation}\label{eq:bloch_freq_def} 
\omega_B=\frac{Fd}{\hbar}\,,
\end{equation}
which manifest themselves both in momentum and real space; this oscillatory phenomenon is well known as Bloch oscillations (BO). The external constant force $F$ breaks the translational symmetry, so that the Bloch states $\spc \bloch^{j}(k)\ket$ are no longer eigenstates of $\C{H}(\alpha=0,\beta=0)$. Nonetheless, we can still use them as a basis set. If interband Landau-Zener (LZ) tunneling can be neglected, one finds the time-evolved wavefunction
\begin{equation}\label{eq:BO2}
\spc \psi(t)\ket =\sum_{j}\frac{d}{2\pi}\int_{\BZ}\mbox{d}k\;G_j(k-Ft/\hbar)\spc
\bloch^j(k)\ket
\end{equation}
where $G_j(k)$ are periodic functions with period $2\pi/d$, which are determined by the initial state at $t=0$. Thus, in case LZ tunneling can be neglected, the system behaves as if the quasi-momentum $k$ evolve{s} in time according to the semiclassical equation of motion
\begin{equation}\label{eq:BO1}
\hbar\,\frac{\mbox{d} k(t)}{\mbox{d}t}= F \,.
\end{equation}
Here, it is implicitly understood that the evolution of quasi-momentum is considered folded within the first Brillouin zone.

The linear increase of the quasi-momentum {results} in a periodic change of the atomic momentum distribution with the angular frequency (\ref{eq:bloch_freq_def}): when the quasi-momentum reaches the boundary of the Bloch band it is Bragg reflected \cite{Peik:1997dx}.

An alternative description of BOs can be given in terms of Wannier-Stark (WS) states $\spc\Psi_{n,j}\ket$, which are quasi-stationary states constructed from the energy band of index $j$ and centered on the lattice site of index $n$ \cite{Gluck:2002p767}. In essence, the $-Fz$ potential breaks the translational symmetry, and the state energies at different sites become mismatched to produce a ladder 
\begin{equation}\label{eq:WSladder}{
\C{E}_{n,j} =
\bar{E}_{j}-n\hspace{1pt}Fd=\bar{E}_{j}-n\hspace{1pt}\hbar\hspace{1pt}\omega_B\,,}
\end{equation}
where the single energy step is determined by the Bloch frequency in Eq.~(\ref{eq:bloch_freq_def}), and $\bar{E}_{j}=d/(2\pi)\hspace{-1pt}\int_{\BZ}\mathrm{d}k\,E_j(k)$ represents the average energy value of the $j$-th band.
 
Bloch tunneling is thus suppressed and the phenomenon of Wannier-Stark localization occurs, though non-zero probability for LZ tunneling introduces a finite lifetime \cite{Gluck:1999ig}.
 
If we consider a state with initial defined wave vector $k_0$ at $t=0$ in the lowest band, its time evolution will be
\begin{multline}\label{eq:BO_WSpicture}
\spc\psi(k_0,t)\ket=\!\!\sum_{n=-\infty}^{\infty}\!e^{-i
  \C{E}_{n,j}t/\hbar}e^{i k_0 n d}\,\spc \Psi_{n,j}\ket=\\
=e^{-i\bar{E}_{j}t/\hbar}\!\!\sum_{n=-\infty}^{\infty}e^{i(k_0 d+\omega_B t)n }\,\spc
\Psi_{n,j}\ket=e^{-i\bar{E}_{j}t/\hbar}\hspace{1pt}\spc\psi(k(t))\ket\,,
\end{multline}
where $k(t)=k_0+\omega_Bt/d$ should be intended modulus the Brillouin zone. The latter result is identical to that of Eq.(\ref{eq:BO2}) previously derived from the semiclassical equation of motion. This second point of view puts in evidence the interferometric character of this phenomenon, where a phase imprint $\omega_Bt$ occurs from site to site. Thus, observing Bloch oscillations requires the initial state to be prepared with \emph{coherences} between separated sites, namely in a superposition of at least two different WS states, e.g., $\spc \Psi_{n,j}\ket$ and $\spc \Psi_{n+1,j}\ket$. Conversely, if the initial state is given by a single WS state, then the dynamics would simply reduce to the accumulation of a trivial global phase. Site-to-site coherence implies in other terms that  the coherence length of the atomic cloud must be longer than the lattice period $d$, which for a thermal gas is of the order of the thermal de Broglie wavelength.
 
Therefore, intuitively a way to enhance the visibility of BO momentum peak is to set a coherent broadening of the initial wave packet. We will come back to this point in Sec.~\ref{sec:inter}, where we show how amplitude and phase modulation of the optical lattices serve to the observation of \BOnew's.
 
\subsection{Driving by nearly resonant amplitude and phase modulation}\label{sec:ModulationDriving}

We study the dynamics resulting from the time-dependent Hamiltonian in (\ref{eq:startH}). We show here that both the amplitude modulation (AM) ($\alpha\neq 0$, $\beta=0$) and phase modulation (PM) ($\alpha=0$, $\beta\neq 0$) of an optical lattice are very flexible tools to steer quantum transport of atomic wave packets in tilted optical lattices, i.e., in the presence of an external constant force $F$. In particular, we are interested in coherent delocalization of matter waves by means of resonant tunneling. Resonant tunneling occurs when the atoms absorb or emit energy quanta $\hbar\s{1} \omega_M$ resonant with the $\ell$-th harmonics of Bloch frequency $\ell\s{1}\hbar\s{1}\omega_B$, while they tunnel upwards or downwards along the optical lattice, respectively. For this reason we will consider nearly resonant modulation frequencies $\omega_M$, i.e. $\omega_M \simeq\ell\,\omega_B$. Energy quanta are exchanged between the atoms and the photons of the modulated optical lattice, suggesting an interpretation of this transport mechanism in terms of photon-assisted tunneling~\cite{Sias:2008p694,Ivanov:2008p37,Kierig2008}. 

In the rest of this work we will consider only physical regimes where interband LZ tunneling is fully negligible. For clarity, we therefore drop the band index $j$, since under this assumption the atomic dynamics remains confined in the same initial lattice band at all times.

In the case of AM driving, we can use the WS basis to rewrite the Hamiltonian in Eq.~(\ref{eq:startH}) in the tight-binding form 
\begin{multline}\raisetag{2.05cm}\label{eq:Ham_start}
\C{H}_{\mathrm{AM}}=\hspace{-6pt}\sum_{n=-\infty}^{+\infty}\hspace{-4pt}-n\hspace{1pt}\hbar\hspace{1pt}\omega_B\spc\Psi_n\ket\bra\Psi_n\spc+\\+
\sum_{n=-\infty}^{+\infty}\hspace{-4pt}\left(\frac{\alpha\hspace{1pt}U_0}{2}\,\C{C}_\ell^{\mathrm{AM}}\hspace{1pt}\sin(\omega_Mt-\phi)\spc \Psi_{n+\ell}\ket\bra \Psi_{n}\spc+\mathrm{h.c.}\right)\,,
\end{multline}
with the coefficients $\C{C}_\ell^{\mathrm{AM}}=\bra \Psi_{n+\ell}\spc \cos(2k_Lz)\spc \Psi_{n}\ket$ representing the real-valued overlap integrals between resonantly-coupled WS states $\spc \Psi_{n}\ket$. In writing the Hamiltonian~(\ref{eq:Ham_start}) we neglected all time-dependent coupling terms, except for those which are nearly resonant. For this reason we disregarded the interband coupling terms, too, as we assume the modulation frequency $\omega_M$ be much smaller than the energy gap between lattice bands. It is possible to express the Hamiltonian in a time-independent form by transforming it into the rotating frame through the operator {$\C{U}_\textrm{RF}=\exp[-i\sum_{n=-\infty}^{+\infty}n\hspace{1pt}(\omega_M/\ell)\hspace{1pt}t\spc \Psi_n\ket\bra \Psi_n\spc]$}. Using the rotating-wave approximation where the fast oscillating terms are neglected, one obtains
\begin{multline}\label{eq:WS_AMham} 
\C{H}'_{\mathrm{AM}}\hspace{-1pt} =\hspace{-8pt}\sum_{n=-\infty}^{+\infty}\hspace{-2pt}\bigg[n\frac{\hbar\hspace{0.5pt}\delta}{\ell} \spc
\Psi_{n}\ket\bra\Psi_{n}\spc\s{1}+\\[1mm]
+\hspace{-2pt}\left(\hspace{-2pt}e^{i\hspace{0.5pt}(\pi/2-\phi)}
\frac{\C{J}_\ell^{\mathrm{AM}}}{2}\spc \Psi_{n+\ell}\ket \bra \Psi_{n}\spc +\mathrm{h.c.}\hspace{-1pt}\right)\s{-2}\bigg]
\end{multline}
where {$\delta=\omega_M-\ell\hspace{1pt}\omega_B$} is the detuning, the prime refer to the transformed reference frame, and the complex-valued coefficients $\C{J}_\ell^{\mathrm{AM}}$ represent the tunneling rates
\begin{equation} \label{eq:tunn_rates} 
\C{J}_\ell^{\mathrm{AM}}=\frac{\alpha\s{1}U_0}{2}\,\bra \Psi_{n+\ell}\spc \cos(2k_Lz)\spc \Psi_{n}\ket\, .
\end{equation}
Both parameters $\alpha$ and $\phi$ can be precisely controlled in an experiment, offering great flexibility in steering atomic transport. For instance, a simple shift by $\pi$ of the phase $\phi$ allows the tunneling rates $\C{J}_\ell^{\mathrm{AM}}$ to be inverted \cite{Alberti:2010p70}.

It is worth noticing that the coefficient $\C{C}_\ell^{\mathrm{AM}}$ can be viewed as the Franck-Condon factor of a stimulated two-photon Raman transition \cite{Deb:2007p1285}, leading to the interpretation of a two-photon Raman process underling the resonant tunneling among intraband WS states~\cite{PhysRevLett.106.213002}. 

In the case of PM driving, following a procedure similar to the one used for AM modulation, one can express the Hamiltonian in the tight-binding form 
\begin{multline}\label{eq:Hpm_start}
	\C{H}_{\mathrm{PM}}=\hspace{-6pt}\sum_{n=-\infty}^{+\infty}\hspace{-6pt}-n\left[\hbar\hspace{1pt}\omega_B+\beta\hspace{1pt}d\hspace{1pt}\sin(\omega_Mt-\phi)\right]\spc\Psi_n\ket\bra\Psi_n\spc\hspace{2pt}+\\
-\hspace{-4pt}\sum_{n=-\infty}^{+\infty}\hspace{-4pt}
\left[\beta\hspace{1pt}d\hspace{2pt}\C{C}_\ell^{\mathrm{PM}}\hspace{1pt}\sin(\omega_Mt-\phi)\spc \Psi_{n+\ell}\ket\bra \Psi_{n}\spc+\mathrm{h.c.}\right]\,,
\end{multline}
where the coefficients $\C{C}_\ell^{\mathrm{PM}}=\bra \Psi_{n+\ell}\spc z \spc \Psi_{n}\ket/d$ represent the real-valued overlap integrals between resonantly-coupled WS states. Similarly to the approximation made in the AM case, in Eq.~(\ref{eq:Hpm_start}) we disregarded all off-resonance time-dependent coupling terms between different WS states; in addition, it is worth underlining that the diagonal time-dependent terms are here kept, in contrast to the AM case, as they exhibit a spatial dependence on the lattice index $n$. Applying the transformation into the rotating frame
 $\C{U}_\mathrm{RF}=\exp(-i\sum_{n=-\infty}^{+\infty}\hspace{-1pt}n\hspace{1pt}\Lambda(t)\hspace{1pt}\spc \Psi_n\ket\bra \Psi_n\spc)$
	with $\Lambda(t)=(\omega_M/\ell)\hspace{1pt} t-\beta \hspace{1pt} d/(\hbar\hspace{1pt}\omega_M) \cos(\omega_Mt-\phi)$ \cite{Drese:1997ky,Gluck:1999,Thommen:2004p6}, the Hamiltonian takes the time-independent form
\begin{multline}\label{eq:Hpm} 
\C{H}'_{\mathrm{PM}}\hspace{-1pt} =\hspace{-8pt}\sum_{n=-\infty}^{+\infty}\hspace{-3pt}\bigg[n\frac{\hbar\hspace{0.5pt}\delta}{\ell} \spc
\Psi_{n}\ket\bra\Psi_{n}\spc\s{1}+\\[1mm]+\hspace{-2pt}\left(\hspace{-2pt}
e^{i\hspace{0.5pt}(\pi/2-\phi)}\frac{\C{J}_\ell^{\mathrm{PM}}}{2}\spc \Psi_{n+\ell}\ket \bra \Psi_{n}\spc+\mathrm{h.c.}\hspace{-1pt}\right)\s{-2}\bigg]
\end{multline}
where the tunneling rates $\C{J}_\ell^{\mathrm{PM}}$ are defined as
\begin{equation} \label{eq:TunnRatePM} 
\C{J}_\ell^{\mathrm{PM}}=2\,\varepsilon_\ell\hspace{1pt}\frac{\omega_M}{\ell\hspace{0.8pt}\omega_B}\,J_1\hspace{-2pt}\left(\frac{\beta\hspace{0.3pt}d\hspace{1pt}\ell}{\hbar\hspace{0.5pt}\omega_M}\right)
\end{equation}
with $J_1$ the first-order Bessel function of the first kind, and $\varepsilon_\ell$ the $\ell$-th Fourier harmonic of the energy band $E(k)$
\begin{equation}\label{eq:epsilon_coeff}
	\varepsilon_\ell=\frac{d}{2\pi}\int_{\BZ}\hspace{-3pt}\mathrm{d}k\,e^{-i\hspace{0.5pt}\ell\hspace{0.5pt}k\hspace{0.5pt}d}\,E(k)\,.
\end{equation}
The quantity $2\,\varepsilon_\ell$ is well known for being the tunneling rate between states lying $\ell$ sites apart in a pure lattice potential, i.e. in the absence of any modulation or external force. A detailed derivation of the Hamiltonian in Eq.~(\ref{eq:Hpm}) is given in Appendix~\ref{appendix:phasemod}.

It is far evident the similarity between (\ref{eq:WS_AMham}) and (\ref{eq:Hpm}), the only and significative difference being given by the actual expression of the tunneling rates (\ref{eq:tunn_rates}) and (\ref{eq:TunnRatePM}). In AM case the tunneling rate is linear in the driving amplitude $\alpha$, while in PM case this occurs only for $\beta d\ll\hbar\hspace{0.8pt}\omega_B$, when $J_1({\beta\hspace{0.3pt}d\hspace{1pt}\ell}/({\hbar\hspace{0.8pt}\omega_M}))\approx{\beta\hspace{0.3pt}d\hspace{1pt}\ell}/(2\hspace{0.8pt}{\hbar\hspace{0.8pt}\omega_M})$. Besides, in the strong-driving regime defined by large values of its argument, $J_1$ vanishes at specific values of the argument and the tunneling rate is dramatically suppressed leading to dynamic localization~\cite{Dunlap:1988p849,Drese:1997ky,Sias:2008p694}.

In both cases, it is demonstrated that the stationary states of (\ref{eq:WS_AMham}) and (\ref{eq:Hpm}) at resonance (i.e, $\delta=0$) are delocalized Bloch states with their energies forming a band 
\begin{equation} \label{eq:energy}
E_\mathrm{mod}(k)=\C{J}_\ell\sin(\ell\hspace{0.5pt} k\hspace{0.5pt}d+\phi)\,,
\end{equation}
where $\C{J}_\ell$ is given, respectively, by $\C{J}^{\mathrm{PM}}_\ell$ or $\C{J}^{\mathrm{AM}}_\ell$ \cite{Ivanov:2008p37,Alberti:2010p70}.

\subsection{State evolution}
\subsubsection{In real space}\label{sec:AmPmMath}

We investigate the time evolution in real space of an initially localized state under the action of AM and PM driving. The transport dynamics is governed by the Hamiltonians in Eqs.\ (\ref{eq:WS_AMham}) and (\ref{eq:Hpm}), respectively, which are similar in their form.

For $\delta\neq0$, the translation invariance is broken and the Hamiltonians exhibit a strict analogy with the Hamiltonian of a static lattice in the presence of an effective homogenous force of magnitude $F_\delta$ defined by $\delta=F_\delta\s{1}\ell\s{1}d/\hbar$. This suggests the physical picture in which the nearly-resonant driving reduces much the ``tilt'' of the lattice potential stemming from the force $F$, but not completely: a small detuning $\delta$ effectively corresponds to a small external force $F_\delta$. One thus expects that the transport dynamics is characterized by Bloch oscillations with time period $2\pi/\delta$. Indeed, this was first observed through macroscopic oscillations of atomic wave packets \cite{Alberti:2009p45}, and later studied with non-interacting Bose-Einstein condensates (BECs) \cite{Haller:2010hx}; it also led to a generalization of Bloch's acceleration theorem for periodically driven systems \cite{2011Arlinghaus}. In the other case with $\delta=0$, the system is invariant under discrete translations, and coherent delocalization occurs, with the atomic wave packets spreading ballistically in time according to the dispersion relation in Eq.\ (\ref{eq:energy}) \cite{Ivanov:2008p37}.

The eigenstates of the Hamiltonians are the Wannier-Stark states
\begin{equation}\label{eq:WSdelta}
	\spc\Phi_n(\delta)\ket= \hspace{-6pt}\sum_{m=-\infty}^{\infty}\hspace{-5pt}e^{i\s{0.6}m\s{0.7}(\pi/2-\phi)}\s{1}J_{-m}\s{-2}\left(\s{-1}\frac{\C{J}_\ell}{F_\delta\s{0.7}d}\s{-1}\right) \spc \Psi_{n+m\s{0.5}\ell}\ket
\end{equation}
corresponding to a force $F_\delta$, with $J_m$ the $m$-th order Bessel function of the first kind \cite{Gluck:2002p767}. These states are centered on the $n$-th lattice site and extend over a number of sites of the order of the argument of the Bessel functions, ${\C{J}_\ell}/({F_\delta\s{0.7}d})$. The set of energies associated to these states forms a Wannier-Stark ladder with $E_n(\delta)=n\s{0.5}\hbar\s{0.3}\delta/\ell$. Hence we obtain the time evolution operator in the rotating frame
\begin{equation}\label{eq:U_t}
\C{U}_T(t) = \exp\left(\s{-4}-i\s{-4}\sum_{n=-\infty}^{\infty}\s{-4}n\s{0.5}\delta\s{0.4}t/\ell\,\spc \Phi_n(\delta)\ket\bra\Phi_n(\delta)\spc\right).
\end{equation}
By applying it to the initial state $\spc \Psi_n\ket$ we get the time-evolved state 
\begin{multline}\label{eq:WS_time}
\spc \Psi_n(t)\ket=e^{-i\s{0.3}n\s{0.3}\delta\s{0.3}t/\ell} \s{-5}\sum_{m=-\infty}^{\infty}\s{-5}e^{-i\s{0.3}m\s{0.3}\phi}e^{-i\s{0.3}m\s{0.3}\delta\s{0.3}t/2}\s{1}\times\\
\times J_m\bigg(\frac{\sin(\delta\s{0.3}t/2)}{\delta\s{0.3}t/2}\,\frac{|\C{J}_\ell|\s{0.3}t}{\hbar}\bigg)\,\spc \Psi_{n+m\s{0.3}\ell}\ket
\end{multline}
at time $t$, with the intermediate steps reported in Appendix\ \ref{appendix:timeevoloper}. The superposition principle allows the evolution of any initial distribution to be calculated, as well.

One can use Eq.\ (\ref{eq:WS_time}) to {calculate} the spatial broadening of a matter-wave packet initially localized at the $n$-th site as a function of time, $\sigma_{\Psi_n}^2(t)\equiv{\bra\Psi_n(t)\spc (z-n\s{0.3}d)^2\spc\Psi_n(t)\ket}$. Considering the expression in Eq.\ (\ref{eq:WS_time}), one expects the quantity $\sigma_{\Psi_n}(t)/d$ to be of order of the argument of the Bessel functions. This is, in fact, confirmed by carrying out the detailed calculation, which produces at large $t$ the asymptotic ballistic expansion
\begin{equation}\label{eq:ResTunnAA}
\sigma_{\Psi_n}^2(t) \approx v_{\ell}^2\s{0.8}t^2\,\sinc\s{-1}
  \left(\delta/\GammaF\right)^2,
\end{equation}
with $v_{\ell}=\hbar^{-1}\ell\s{0.5}d\s{0.7}|\C{J}_\ell|/\sqrt{2}$ the speed of broadening at resonance, $\GammaF=2\pi\times(\pi t)^{-1}$ the Fourier-limited linewidth of the resonance spectrum, and the $\sinc$ function defined as $\sinc(x)=\sin(x)/x$. Appendix\ \ref{appendix:sigma} reports a derivation of this expression for the resonant case ($\delta=0$).  We omitted showing in Eq.~(\ref{eq:ResTunnAA}) the terms corresponding to the size of the initial state $\spc \Psi_n\ket$ and a quivering motion exhibiting fast, tiny oscillations on a timescale of the order of $2\pi/\omega_M$; in fact, these terms are of marginal significance, as they depend specifically on the given initial state. The complete expression is however provided in Appendix\ \ref{appendix:sigma}.

Thus, from the expression~(\ref{eq:ResTunnAA}) we see that resonant {driving} has the effect of coherently delocaliz{ing} atomic wave packets over sites which are distant $\ell d$ apart; conversely, nearly-resonant driving yields spatial oscillations with period $2\pi/\delta$ and maximal extent $2\s{0.3}v_\ell/\delta$. The prediction in Eq.\ (\ref{eq:ResTunnAA}) for $\sigma_{\Psi_n}(t)$ allows the measurement of Bloch frequency by directly determining the resonant frequency $\ell\s{0.3}\omega_B$ through \emph{in-situ} observations of the spatial distribution of the atoms in the lattice. In reality, one has to consider that the atomic cloud is characterized by an initial finite size, $\sigma_0$, which produces a spatial distribution that is the convolution of the two distributions: the variance, corrected for the atomic cloud's size, becomes 

\begin{equation}\label{eq:ResTunn}
\sigma(t) =\sqrt{\sigma_0^2+ v_{\ell}^2\s{0.8}t^2\,\sinc
  \left(\frac{\omega_M-\ell\s{0.3}\omega_B}{\Gamma}\right)^2}.
\end{equation}   
  
This expression depends on four free-parameters, which can be determined by fitting the experimental data: the starting width of the atomic cloud $\sigma_0$, the maximal extent at resonance $v_\ell\s{0.3}t$, the resonance linewidth $\Gamma$, and the Bloch frequency $\omega_B$. The experimental validation of this formula will be analyzed in Sec.~\ref{sec:SensitivityGeneral}.

\subsubsection{In momentum space}

Studying nearly-resonant driving in momentum space brings an alternative viewpoint, which puts in the foreground mainly the aspects related to quantum transport. One in fact expects that at resonance the eigenstates of Eq.\ (\ref{eq:U_t}) are delocalized Bloch states $\spc k\ket$ with wavevector $k$, and the transport dynamics is determined by the energy dispersion relation $E_\mathrm{mod}(k)$ introduced in Eq.\ (\ref{eq:energy}). Conversely, when off resonance the eigenstates are the Wannier-Stark states $\spc\Phi_n(\delta)\ket$ given in Eq.\ (\ref{eq:WSdelta}), and one expects that the transport-dynamics is characterized by Bloch-like oscillations: the quasi-momentum increases linearly in time as \mbox{$\hbar\s{1}\mathrm{d}k/\mathrm{d}t=-F_\delta\s{1}t$}, with the understanding that $k$ is taken modulus the Brillouin zone.

This can be directly seen by projecting the time-evolved state $\Psi_n (t)\ket$ on quasi-momentum states,
\begin{multline}\label{eq:dimostr}
\bra k\spc \Psi_n (t)\ket =\bra k\spc \Psi_n\ket\s{2}e^{-in\delta t/ \ell}\s{1}\times\\
\times  \exp\s{-3}\left(\s{-3}-\frac{i}{\hbar}\s{-1}\int_{0}^{t}\s{-3}\mathrm{d}t'\s{2} E_\mathrm{mod}(k+F_\delta t'/\hbar)\right),
\end{multline}
where we made use of Bloch's theorem which provides translational symmetry to the WS states
\begin{equation}\label{eq:tran_symm}
\bra k\spc \Psi_{n+m}\ket=e^{-i\s{0.3}m\s{0.3}k\s{0.3}d}\s{0.6}\bra k\spc \Psi_{n}\ket
\end{equation}
and the Jacobi-Anger expansion to derive it.

This result allows us to interpret an optical lattice that is dynamically driven by AM or PM modulation as a tilted stationary lattice whose parameters can be tailored at will. The tilt is controlled though the detuning $\delta$. The periodicity of $E_\mathrm{mod}(k)$ as a function of $k$ shows that the size of the Brillouin zone is shrunk by the order of the driving harmonic $\ell$. This is rigorously proven in the next section. The overall phase of the effective energy band $\phi$ can be tuned through the phase of the time-dependent external driving. The width of the energy band $\C{J}_\ell$ can be manipulated as well by tuning the driving strength $\alpha$ and $\beta$. In the PM case the width depends also on the ratio $\beta\s{0.3} \ell\s{0.3} d/(\hbar\omega_M)$, due to the mechanism of dynamic localization for strong driving.

Besides, the result (\ref{eq:dimostr}) implies that the momentum spectrum of an initially localized WS state evolves in time in the presence of external driving by only acquiring a phase, which depends on the time and on the wave vector $k$. Then, the squared modulus of the momentum spectrum remains constant
\begin{equation}
\label{eq:dubfirub93ionrocbnr}
\spc\bra k \spc\Psi_n(t)\ket\spc^2=\spc\bra k \spc\Psi_0\ket\spc^2,
\end{equation}
implying that only the probability amplitudes and not the probability distribution in the momentum space evolve in time.

However, dynamical driving can be used to extend spatial \emph{coherences} of wave packets, i.e. delocalizing them over several lattice sites, leading to transient narrow distributions in time-of-flight (TOF) experiments. Transient modification of TOF density distributions is at the basis of \BOnew{} mechanism, and it is discussed in Sec.\ \ref{sec:inter}; in static optical lattices, transient interference phenomena have been recently investigated using BECs~\cite{Gerbier:2008p18}.

\subsection{Transport resonance spectrum}\label{sec:TRS}

We want to study the problem of quantum transport in driven optical lattices from the point of view of space and time symmetries. In particular, we are interested in determining the exact positions of resonances without making any approximation of the original Hamiltonian in Eq.\ (\ref{eq:startH}). This is of special interest for the application of AM or PM driven quantum transport to determine the Bloch frequency $\omega_B$ in Eq.\ (\ref{eq:bloch_freq_def}), thus performing a precise and accurate measurement of the local force F.

It is convenient to define $\C{H}_0$ as the Hamiltonian in the absence of modulation, i.e., for $\alpha=0$, $\beta=0$, and $\C{H}_1$ as the driving Hamiltonian $\C{H}_1=\C{H}-\C{H}_0$. Regarding space symmetries, we have that 
$\C{D}_m\C{H}_0\C{D}_m^\dag=\C{H}_0+m\hspace{1pt}\hbar\hspace{0.3pt}\omega_B$ and $\C{D}_m\C{H}_1\C{D}_m^\dag=\C{H}_1+m\hspace{1pt} \beta\hspace{1pt}d\hspace{1pt} f(t)$
with $\C{D}_m$ the unitary operator shifting the space coordinate by an integer number $m$ of lattice sites. Regarding time symmetries, we have that $\C{H}_0$ does not depend on time, while $\C{H}_1$ is invariant under time shifts by an integer number of $\tau_M=2\pi/\omega_M$.

The invariance under discrete-time shifts, which occurs for periodic drivings, allows the use of the Floquet theory \cite{Grifoni:1998tk} in order to express the time evolution operator $\C{U}_T(t)$ at the time $t=\kappa\,\tau_M$ as
\begin{equation}
\C{U}_T(\kappa\hspace{1pt}\tau_M)=\C{U}_T(\tau_M)^\kappa
\end{equation}
with $\kappa$ an integer value. The Floquet quasi-energies $E_i$ are determined, up to integer multiples of $2\pi\hspace{0.5pt}\tau_M/\hbar$, by means of the eigenstates $\exp(-i\hspace{1pt}E_i\hspace{1pt}\tau_M/\hbar)$ of $\C{U}_T(\tau_M)$: transport resonances manifest themselves when the corresponding eigenstates of $\C{U}_T(\tau_M)$ are delocalized states, leading to coherent delocalization of any initially-localized wave packet. In addition, as a result of Bloch's theorem, we have that the eigenstates are indeed delocalized when
$\C{D}_m\hspace{1pt}\C{U}_T(\tau_M)\C{D}_m^\dag=\C{U}_T(\tau_M)$, i.e., when $\C{U}_T(\tau_M)$ is invariant under translations of an integer number of lattice sites.

By making use of the interaction picture, the evolution operator can be written as
\begin{equation}
	\C{U}_T(\tau_M)=\exp(-i\hspace{1pt}\C{H}_0\tau_M/\hbar)\hspace{1pt}\,T\hspace{-0.5pt}\exp
	\left(\hspace{-1.5pt}-i\hspace{-1.5pt}\int_0^{\tau_M}\hspace{-7pt}\mathrm{d}t\hspace{2pt}\tilde{\C{H}}_1/\hbar\hspace{-0.5pt}\right),
\end{equation}
where $T\hspace{-0.5pt}\exp$ represents the time-ordered exponential and $\tilde{\C{H}}_1=\exp(i\hspace{1pt}\C{H}_0\tau_M/\hbar)\hspace{2pt}\C{H}_1\exp(-i\hspace{1pt}\C{H}_0\tau_M/\hbar)$ is the transformed Hamiltonian. Owing to the space symmetry mentioned above, it follows that the first term transforms as $\C{D}_m\exp(-i\hspace{1pt}\C{H}_0\tau_M/\hbar)\hspace{1pt}\C{D}_m^\dag=\exp(-i\hspace{1pt}\C{H}_0\tau_M/\hbar)\hspace{1pt}\exp(-i\hspace{1pt}m\hspace{1pt}\omega_B\hspace{0.5pt}\tau_M)$, while the second one remains invariant
\begin{multline}
	\C{D}_m\hspace{1.5pt}T\hspace{-0.5pt}\exp
	\left(\hspace{-1.5pt}-i\hspace{-1.5pt}\int_0^{\tau_M}\hspace{-7pt}\mathrm{d}t\hspace{2pt}\tilde{\C{H}}_1/\hbar\hspace{-0.5pt}\right)\C{D}_m^\dag=\\=\exp\left(\hspace{-2.5pt}-i\hspace{-1.5pt}\int_0^{\tau_M}\hspace{-8pt}\mathrm{d}t\hspace{3pt}m\hspace{1pt} \beta\hspace{1pt}d\hspace{1pt} f(t)/\hbar\hspace{-0.5pt}\right)T\hspace{-0.5pt}\exp
	\left(\hspace{-2.5pt}-i\hspace{-1.5pt}\int_0^{\tau_M}\hspace{-8pt}\mathrm{d}t\hspace{2pt}\tilde{\C{H}}_1/\hbar\hspace{-0.5pt}\right),
\end{multline}
where the identity $\int_0^{\tau_M}\hspace{-2pt}\mathrm{d}t\hspace{1pt} f(t)=0$ is enforced by the definition of $\tau_M$. It results finally that
\begin{equation}
	\C{D}_m\hspace{1pt}\C{U}_T(\tau_M)\C{D}_m^\dag=\C{U}_T(\tau_M)\hspace{0.5pt}\exp(-i\hspace{1pt}m\hspace{1pt}\omega_B\hspace{0.5pt}\tau_M),
\end{equation}
implying that $[\hspace{2pt}\C{U}_T(\tau_M), D_m]=0$ for $\omega_M=\omega_B/q$ with $q$ an integer number. More generally, if we consider the symmetry group characterized by discrete translations of $\ell\,m$ lattice sites, with $\ell$ and $m$  integer numbers, it follows that $[\hspace{2pt}\C{U}_T(\tau_M), D_{\ell\hspace{0.3pt}m}]=0$ for $\omega_M=\ell\hspace{1pt}\omega_B/q$, proving that transport resonances, i.e. coherent delocalization of atomic wave packets, occur at integer (and sub integer) harmonics of the Bloch frequency $\omega_B$.

In addition, this result proves that, when modulating at integer harmonics $\ell$, the translation symmetry group is the one determined by a unit step of $\ell$ lattice sites, indicating that the first Brillouin zone which describes the eigenstates of $\C{U}_T(\tau_M)$ extends over the interval $[-\pi/(\ell\hspace{0.5pt}d),\pi/(\ell\hspace{0.5pt}d)]$, thus showing indeed a shrinking by an integer factor $\ell$ of the Brillouin zone of the original lattice.

These findings can be straightforwardly generalized to an arbitrary periodic driving $f(t)$ with period $2\pi/\omega_M$. It results that for AM driving the location of transport resonances remains unchanged, while for PM driving one should consider that a non-zero average value of the driving force $F'=\int_0^{\tau_M}\hspace{-2pt}\mathrm{d}t\hspace{1pt}\beta\hspace{0.5pt}f(t)/\tau_M$ produces an effective Bloch frequency shift $F'd/\hbar$ with respect to $\omega_B$ defined in Eq.\ \ref{eq:bloch_freq_def}. This generalization to arbitrary periodic drivings is of clear interest for measuring $\omega_B$ with high precision and accuracy: measurements obtained by AM driving are insensitive to the presence of spurious higher and sub harmonics of $\omega_M$ in the modulation signal $f(t)$.
 
\subsection{Delocalization-enhanced Bloch oscillations}\label{sec:inter} 

In Sec.~\ref{sec:AmPmMath} we showed how resonant driving of the lattice potential establishes coherences among WS states as a result of the broadening of atomic wave packets during the modulation time. In the case of a finite time of free evolution after releasing from the optical lattice, this implies interference of matter waves initially located at different lattice sites. This can be put into evidence by means of the following three-steps experiment: we apply the (AM or PM) modulation $f(t)=\sin(\omega_Bt)$  for a certain time $t_{mod}$, then hold the wave packets in the static lattice for $t_{hold}$, and finally release the wave packets in time of flight $t_{tof}$. For an optimal choice of $t_{mod}$ and $t_{tof}$ it is possible to enhance the interference effect responsible for BOs and its visibility. 

To understand the \BOnew~mechanism, we thus start with an initially localized state $\spc \psi_0\ket$ centered around a site $n=0$. The results could be easily generalized to any initial site $n$ by a simple translation. The initial state at $t=0$ can be written as in (\ref{eq:BO2}). If the thermal energy $k_BT$ is larger than the recoil energy $E_R$, $G(k)$ is smooth and flat over the whole Brillouin zone. In the first step, after a modulation at the $\ell$-th {harmonic} for a time $t_{mod}$, the states $\spc \bloch(k)\ket$ thus evolve by means of the operator $\C{U}(t) = \exp(-i\,E_{mod}(k)\,t/\hbar)$:

\begin{equation}
\label{eq:dcidbufjd}
\spc \bloch(k,t_{mod})\ket =e^{-i\,A\,\sin(\ell kd)}\spc \bloch(k)\ket
\end{equation}
where the quantity $A\equiv\C{J}_{\ell}\,t_{mod}/\hbar$ multiplied by $\ell$ represents the spatial extent of the broadened wave function expressed in units of the lattice constant $d$.

During the hold time $t_{hold}$ in the static lattice, the external force shifts $k$ according to {$k\rightarrow k+\omega_Bt_{hold}/d$} and the wave function changes into

\begin{multline}\label{eq:ifncrdfn}
\spc\psi(t_{mod}+t_{hold})\ket=\\= \int_{\BZ}
\mbox{d}k\ G(k)\,e^{-i\,A\,\sin(\ell kd)}\spc
\bloch(k+\omega_Bt_{hold}/d)\ket
\end{multline}

The result (\ref{eq:ifncrdfn}) can be alternatively stated as:

\begin{multline}\label{eq:ifncrdfndfdd}
\spc\psi(t_{mod}+t_{hold})\ket= \int_{\BZ}
\mbox{d}k\,G(k-\omega_Bt_{hold}/d)\times\\
e^{-iA\sin(\ell kd-2\pi\ell  t_{hold}/\tau_B)}\spc \bloch(k)\ket
\end{multline}
where the argument of the $G$ function has to be taken modulus the Brillouin zone, and $\tau_B$ is the Bloch period. Switching off the lattice maps the Bloch states into free particle states. This procedure can occur suddenly or adiabatically. While in the first case complex {interference effects involving higher energy bands} are expected to appear, we want here to catch the basic concept in the simplest case of adiabatic release. After a time of flight $t_{tof}$, the probability amplitude to find the atom along the optical lattice direction is:

\begin{multline}\label{eq:IntPeak}
\bra z\spc\psi(t_{mod}+t_{hold}+t_{tof})\ket= \int_{\BZ}
\mbox{d}k\,G(k-\omega_B\,t_{hold}/d)\,\times\\\times e^{-i\,A\,\sin(\ell kd-2\pi\ell
  t_{hold}/\tau_B)}{}e^{-i \,\hbar
  k^2t_{tof}/{2M}}e^{ikz}
\end{multline}

\begin{figure}[tb]
\begin{center}
\includegraphics[width=0.5\textwidth]{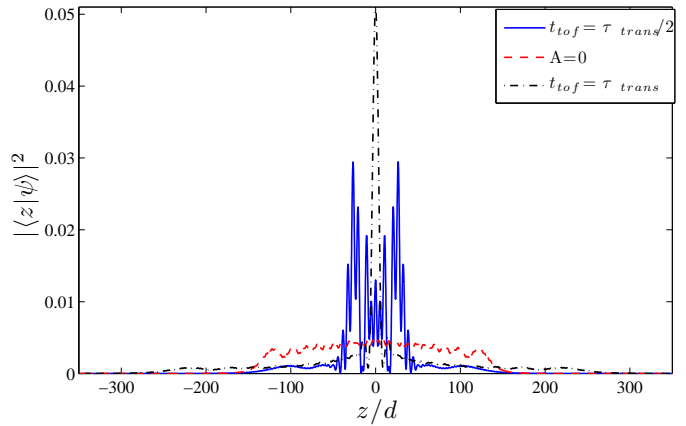}
\caption{Enhancement of the interference peak during time-of-flight detection of Bloch oscillations for a thermal wave packet {($k_BT/E_R \sim 3$)} by means of a resonant modulation burst ($\ell=1$). The solid (blue) line represents the single atom wave function after the three steps BO measurement with a broadening factor $A = 30$ lattice sites and $t_{tof} = M\,A\,d^2/(2\hbar)$ which exhibits a narrow feature at the center of the wave packet with respect to the  wave packet without initial broadening (blue dashed line). The third (black dash-dotted) line shows the wave function at $t_{tof} =  M\,A\,d^2/(\hbar)=\tau_{trans}$ which shows the narrowest peak.}
\label{fig:Sim1}
\end{center}
\end{figure}

The integral in (\ref{eq:IntPeak}) determines the shape and position of the wave packet. In particular we are interested in the interference peak {which is produced as a caustic effect because of} dispersionless evolution {at the interference points}. Since we supposed a flat distribution $G(k)$,  the interference peak is explained in terms of a vanishing second derivative of the phase in the exponent {$g(k)=A\,\sin(\ell kd-2\pi\ell \,t_{hold}/\tau_B)+\hbar\, k^2\,t_{tof}/(2M)$}. We also require a vanishing third derivative to make the region with $g''(k)=0$ as large as possible. The two conditions are satisfied when $t_{tof}=\tau_{trans}$ where

\begin{equation}
\label{eq:dbfueirubw}
\tau_{trans} \equiv M\,A\,\ell^2d^2/\hbar
\end{equation}
is the transient time after which the visibility of the interference peak is maximal. Fig. \ref{fig:Sim1} clearly displays the emergence of a dispersionless peak in the wave packet as the time of flight approaches $\tau_{trans}$.

The transient time $\tau_{trans}$ (\ref{eq:dbfueirubw}) has a clear physical explanation. Since $k_BT>E_R$, the wave packet is essentially localized in a single lattice site, from where it broadens in time of flight with a speed $\hbar/(M\,\ell\,d)$. Only after a time $\tau_{trans}$ the time-of-flight expansion is equal to the extent of the broadened wave packets after modulation which is $\ell\,d\,A$. Picturing the wave packet as a superposition of different waves initially located in different lattice sites, the maximal interference occurs when all the broadening waves in time of flight meet each other, which occurs at the transient time $\tau_{trans}$. The \BOnew~technique can be applied for measurements with superior sensitivity to the effect of external forces. The details of the experimental results are presented in Sec.\ref{sec:SensitivityGeneral}.

\subsection{Decoherence in momentum space and resonant tunneling}\label{sec:decoherence}

Since both BO and WS resonant tunneling are coherent quantum phenomena, it is important to investigate the effects of decoherence. Basic decoherence mechanisms are spontaneous emission, atom-atom interactions~\cite{Kolovsky:2004p22}, lattice random recoils~\cite{Kolovsky:2002p1611} and environmental scattering (i.e., off-resonance scattering by incoherent light or background gas collisions).

While both spontaneous emission and environmental scattering do not affect the dynamics of an ultra-cold gases in a high vacuum system and in their ground state, it is possible to evaluate the effect of atom or photon scattering. They can in fact have different strengths, leading to a collapse of the atomic wave function in a single site, thus leading to a decrease of the BO visibility in momentum space or sequential tunneling in position space, even in the presence of a modulation.

The effect of a decoherence process in momentum space can be understood in terms of superposition of WS states. Starting from (\ref{eq:BO_WSpicture}) (with $\alpha=\beta=0$) and exploiting the translational symmetry of WS states (\ref{eq:tran_symm}), we can rewrite a general expression for the atomic wavefunction in the momentum space~\cite{Witthaut:2005p14}

\begin{multline}\label{eq:BO_dec1}
\bra k\spc\psi(t)\ket=\bra k\spc\Psi_0(k)\ket\cdot\sum_{n=-\infty}^{\infty}\!c_ne^{-i2\pi n(k+Ft/(d\hbar))}\\
 = \bra k\spc\Psi_0(k)\ket\, C(k+Ft/(d\hbar))\, .
\end{multline}

Thus the wave function is the result of the product of a single WS momentum distribution and an envelope function $C(k)$, which is a comb-like distribution periodic both in time and in momentum space. The width of the momentum peaks  $\Delta k$ is inversely proportional to the number $N$ of the superimposed WS states, as we already showed in the previous section. Any scattering event like photon recoil or atom-atom collisions yields a decay of the WS states superposition, while conversely a diffusion-like behavior of the wave packet dispersion. Thus we expect that the momentum peak width evolves with time as 

\begin{equation}\label{eq:BO_dec2}
 \Delta k^2 (t)\simeq \Delta k(0)^2(1+\gamma\, t)
\end{equation}
where $\gamma$ is the rate of the scattering process. For an exact treatment of the dynamics of the wave-packet, and thus of the peak width $\Delta k$, it is necessary to solve the stochastic Liouville equation \cite{Kenkre:1985p1864}, where a localization operator is introduced and a \emph{master equation}  for the evolution of  the density matrix $\rho$ is provided. The prediction for the momentum peak broadening in~(\ref{eq:BO_dec2}) is equivalent to say that a random Gaussian process heats the atomic gas until a uniform distribution of the atoms over the Brillouin zone is obtained. This random walk in the momentum space is proportional to the time $t$ and the density of scatterers (the atoms or the lattice photons).

In the case of BO experiment, we must also consider the presence of the excited bands which introduce additional source of losses. In fact for shallow traps ($U_0\gtrsim E_R$) collisions can result in the transition of the atom from the lowest band into the upper bands which are only weakly bounded. This means that only collisions with small exchanges of momentum can effectively contribute to the broadening of $\Delta k$ and hence $\gamma_{exp} < \gamma$~\cite{Kolovsky:2002p1611}. The lattice-recoil effect plays a smaller role since atom-lattice scattering energy exchange excites the atoms into the second band (the energy gap between the first and the second band is typically $E_G\sim E_R$) and thus does not contribute to the BO signal.

In the case of resonant driving applied to the system, the localization process inhibits the coherent coupling among adjacent WS states which is responsible of the resonant tunneling of the initial atom distribution. The decay of the off-diagonal elements of the density matrix causes a sequential tunneling which asymptotically gives, as already stated, $ \sigma^2 (t) \approx D\,t$. The sequential tunneling processes have the effect of broadening the resonance spectrum. In the case of simple localization Lindblad operators $L_{n}=\sqrt{\gamma}\,\spc n\ket\bra n\spc$, the numerical integration of the master equation leads to a Lorentzian-shaped resonant tunneling spectrum with $\gamma$ approximatively the linewidth of the spectrum~\cite{phdAlberti}. The presence of a decoherence process will result into a broadening of the Fourier-limited linewidth nearly by an additive quantity $\gamma$, which is the inverse of the sensor coherence time.

Having presented decoherence from both points of view of momentum and real space, we will discuss quantitatively the impact of decoherence on the final sensitivity in Sec.~\ref{sec:PS_BO}.

\section{Experimental setup and procedures}
\label{sec:ExperimentalSetup}

\begin{figure}[!t]
\begin{center}
 \includegraphics[angle=90,width=1.05\columnwidth]{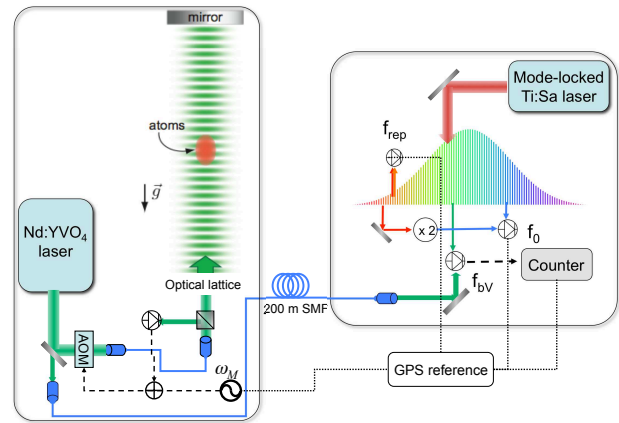}
 \caption{Experimental setup. The output radiation from {a single mode 532~nm (Verdi)} is split and then {focused} on the ultra-cold $^{88}$Sr cloud, while {a pick-off} is sent  through 200 m fiber to the comb lab for frequency measurements. $f_{bV}$: beat-note {frequency} between Verdi and comb; $f_0$: the carrier-envelope offset frequency, $f_{rep}${:} repetition rate frequency; {SMF: single mode fiber; AOM: acousto-optic modulator}.}
\label{fig:ExpSetup}
\end{center}
\end{figure}

{The experimental} setup is based on cooled and trapped $^{88}$Sr atoms \cite{Ferrari06}. The details of the atomic sample preparation was already described elsewhere \cite{Poli05}. Here we limit our description to those basic elements which have a relevance on explaining the results in terms of accuracy and sensitivity of our atomic force sensor. A schematic picture of the experimental setup is shown in Fig.\ref{fig:ExpSetup}.

Atoms from a thermal beam are slowed by a Zeeman slower and trapped in a ``blue'' magneto-optical trap (MOT) operating on the $^1 S_0-^1P_1$ resonance transition at 461 nm. The temperature is further reduced by a second cooling stage in a ``red'' MOT operating on the $^1S_0-^3P_1$ intercombination transition at 689~nm. This produces about $10^6$ atoms at a temperature of 1 $\mu$K. Since the force of gravity is comparable to the force exerted by the red MOT on the atoms, the cloud of trapped atoms assumes a disk-like shape with a vertical size of 27 $\mu$m and a radial size of 180 $\mu$m. The atoms are adiabatically loaded in an optical lattice in 200~$\mu$s. The lattice potential is generated by a single-mode frequency-doubled $\textrm{Nd:YVO}_4$ commercial laser  (Verdi-V5 Coherent, $\lambda_L = 532$~nm) delivering about 1~W on the atoms. The optical lattice is then produced either by retro-reflecting the input beam or by means of a second fiber which delivers half of the available power from the bottom side of the vacuum chamber~\cite{SorrentinoPRA}. A telescope can set the beam waist between 300 and 560 $\mu$m at will.

Typical experimental parameters yield a lattice population of $N_{at} = 5\cdot 10^4$, for an atomic density $n = 6\cdot 10^{11}$ atoms/cm$^3$ and lattice depth $ U_0 = 3 E_R$. If we consider the decoherence due to elastic collisions driven by a scattering length $a_{88} \,=\, -1.6\, a_0$ \cite{PhysRevA.81.051601}, we expect $\gamma_{coll} < 2.2\cdot 10^{-3} $ s$^{-1}$. A high vacuum system allows to exploit the long coherence time allowed by strontium atoms. It consists of a 50 $l/s$ ion pump and a titanium-sublimation pump yielding a background pressure level smaller than $10^{-10}$ Torr. The measured lifetime in the vertical lattice is about 20~s.

We used earth gravity as a test force for the atomic sensor. This is possible by placing the optical axis of our lattice along the direction of $\vec{g}$. The beam is aligned along the gravity axis as follows: the superposition of one beam onto the other is obtained by maximizing the transmission through the fiber of the opposite beam. Then the lattice verticality is made sure by interposing a large glass container with water over the vacuum cell where atoms are trapped. Part of the upwards-directed beam is reflected by this surface with an angle $\theta$. Finally we measure far apart ($D=3\,m$) from the cell the distance $s\simeq\theta D$ between the laser spot formed by this reflection and the second downwards-directed laser beam passing through the water container. This distance $s$  is then reduced to zero  through the fine adjustment of the optical table tilt. This experimental procedure then prevents any possible systematic shift due to both relative and vertical alignment of the two counter-propagating laser beams. The residual systematic uncertainty is described later.

Absorption imaging of the atomic cloud position and distribution is performed either {\it in situ} or by means of time-of-flight (TOF) technique by a CCD camera with a spatial resolution of 4.6~$\mu$m. 

Different procedures can be applied to measure $g$. In the case of PM resonant tunneling, a piezoelectric actuator was attached to the retro-reflecting mirror. The AM technique, instead, employs one double-pass acousto-optic modulator (AOM) which is servo controlled to simultaneously stabilize the lattice depth and provide the modulation frequency $\omega_M$ to the lattice. This modulation is either continuous or with a sequence of modulation bursts depending on the kind of experiment. The former operational method is used to set resonant tunneling, the latter for \BOnew~ or Loschmidt echo (which is not discussed in this work, but detailed in Ref.~\cite{Alberti:2010p70}) experiments. Both the modulation frequency and the experimental timing are referenced to a GPS-stabilized signal, as shown in Fig.\ref{fig:ExpSetup}. 

\begin{figure}[!t]
\begin{center}
\includegraphics[width=0.95\columnwidth]{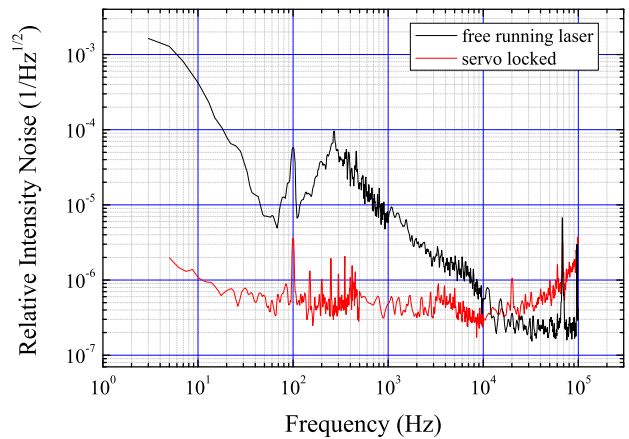}
\caption{Noise spectra of the relative intensity of the lattice laser at 532~nm which is used to produce the optical lattice potential. An intensity stabilization system strongly suppresses the intensity noise of the laser. The comparison of the two spectra shows that the servo lock attenuates the bump of noise originating from the acoustic vibrations by about 30$\div$40 dB.} 
\label{fig:AMservo}
\end{center}
\end{figure}

A servo control system is employed to stabilize the laser intensity, and thus avoiding either slow drifts of the lattice depth $U_0$ and suppressing acoustic vibration-induced amplitude noise. The laser intensity is controlled by changing the radio frequency (RF) signal which drives an AOM. The error signal is then obtained by comparing the intensity measured from a pick-off after the optical fiber and a stable voltage reference. Hence, any misalignment of the fiber injection does not affect the lattice potential depth. Instead fluctuations of the polarization after the optical fiber can also affect the potential depth, mostly in a configuration of two counter-propagating laser beams with respect to retro-reflected single beam. The stabilized intensity noise spectra in Fig.~\ref{fig:AMservo} shows the reduction of the free-running amplitude noise in the acoustic range by about two orders of magnitude, with a servo bandwidth of about 20 kHz. The noise frequencies which are most harmful to the atoms are, in fact, in the range from Hz to several kHz, which effectively coincide with the typical acoustic frequencies. Excitations at these frequencies induce heating of the atoms by exciting the vibrational frequencies of the optical lattice \cite{Savard97}. As shown in Fig.~\ref{fig:AMservo}, we achieve a stabilization of the intensity offset at the relative level of $10^{-6}$.

The servo loop controlling the lattice laser beam intensity also allows to insert the AM driving signal which shares the same frequency reference as the other relevant frequencies. From the recorded FFT spectra of the modulated laser intensity signal the higher harmonics contents have been reduced of more than $60\,\R{dB}$, i.e. below the noise floor level. This reduces the projected Bloch frequency uncertainty at 20 ppb level (see Sec.~\ref{sec:WSsensitivity}). It is evident that the possibility to have an in-loop driving modulation in the AM resonant tunneling and a direct measurement of the driving into the lattice make the AM technique more suitable for high precision measurement with respect to the PM technique employing a mechanical transducer. 

The PM technique is performed by applying the driving signal to the PZT attached to the retro-reflecting mirror. The response of this actuator was characterized by an heterodyne measurement with a Michelson interferometer, which shows a linear dynamical response of the electro-mechanical system with a cut-off frequency of 4 kHz~\cite{phdAlberti}. The voltage applied to the PZT allows to modulate the position the lattice potential by up to 10 lattice sites peak to peak, so that $z_0 \leq 2.5 \lambda_L$. According to (\ref{eq:TunnRatePM}), this means that the non-linear regime showing dynamic localization can occur only at modulation frequencies higher than the 6-th harmonic.

\subsection{Calibration of the lattice frequency}

A precise knowledge of the lattice laser frequency value and its instability is needed for high precision force measurements. We employed a Ti:Sapphire femtosecond frequency comb to obtain the absolute frequency measurements of the lattice laser and to characterize its frequency stability \cite{beverini:79931I}. As shown in the schematic diagram of the experimental setup in Fig.\ref{fig:ExpSetup}, the output radiation from the 200-meter-long fiber is superimposed with radiation from the femtosecond comb (filtered at $\lambda = 532\pm5$ nm) and the beat signal with the corresponding tooth is observed on a fast photodetector.

\begin{figure}[!t]
\begin{center}
 \includegraphics[width=0.98\columnwidth]{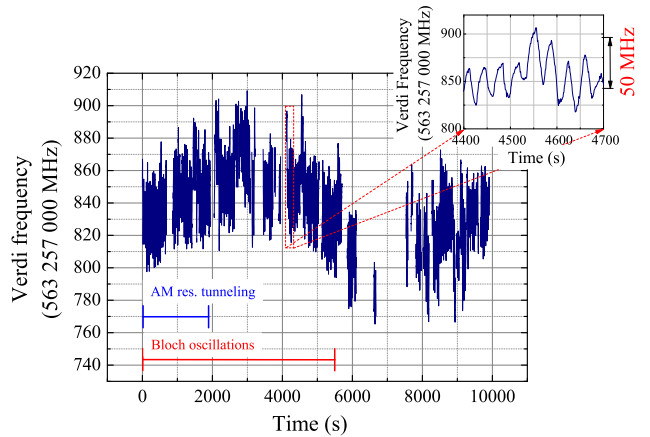}
\caption{Example of frequency counting of Verdi laser for calibration of the 1D $^{88}$Sr optical lattice frequency, reporting for comparison the typical duration of gravity measurements for both techniques. The beat-note signal between lattice laser and the optical comb is used to stabilize the repetition rate of the femtosecond laser. The intervals where no data is reported are caused by the optical frequency comb running out of lock from the lattice signal.}
\label{fig:VerdiCount}
\end{center}
\end{figure}

To perform precise frequency measurement, we lock the repetition rate of the comb by stabilizing this beat-note to a RF-frequency synthesizer and we count the carrier-envelope offset frequency of the optical frequency comb {by self-referencing techniques}. The absolute value of the lattice laser frequency is given by equation $f = n \times f_{rep} \pm f_0 \pm f_{bV}$, where the correct tooth number $n$ is determined by a wavemeter calibrated with a stabilized laser on the 689 nm $^1$S$_0$-$^3$P$_1$ $^{88}$Sr transition, that allows an evaluation of the lattice laser frequency with an uncertainty of the order of 100 MHz.

Fig.\ref{fig:VerdiCount} displays the result of an absolute frequency measurement that lasts for about $10^4$ s. It is interesting to compare dynamics of the lattice laser frequency with typical durations of Bloch frequency measurement both with BO technique and with AM resonant tunneling. From this measurement we can clearly observe periodical fluctuations at two different regimes: the first low frequency fluctuation with typical period of about $10^4$ s and amplitude of 130--140 MHz, and a second regime is a fast oscillation with a shorter period of about 33 s and a typical amplitude of 30--70 MHz (see inset in Fig.\ref{fig:VerdiCount}).

Long term deviation of lattice laser frequency greatly depends on many factors, mainly related to temperature and laser power setting. To increase the long term stability we left the laser continuously operating for about two days at the same power setting. Moreover, a thermal insulation foam is applied to the laser head to increase the passive temperature stability. However, this model of commercial laser can show both quite different absolute frequency (offset by several GHz) and large frequency drift rate at start up (up to 100 MHz in one minute).

Hence, long-lasting measurements of forces by means of the Bloch frequency are affected by lattice frequency uncertainty at the 100 ppb level. Looking at the length of the two force measurements, BO measurements usually requires $5.6\cdot10^3$ s (1.5 hours), while AM resonant tunneling lasts three times less. Then the sensitivity of BO measurements is three times higher, increasing the error for each Bloch frequency determination, though the systematic effect can be equivalent in both techniques.

\section{Measuring external forces: Bloch oscillations vs. driven resonant tunneling}\label{sec:SensitivityGeneral}

\subsection{Effects at resonant driving I: Wannier-Stark localization
  {\it vs.} coherent delocalization}

As discussed in Sec.~\ref{sec:AmPmMath}, atomic wave packets loaded into a driven vertical optical lattice potential exhibit {coherent delocalization} arising from intraband transitions among WS levels. WS intraband transitions are observed by monitoring the \emph{in-situ} wave packet extent. Coherent delocalization sets in for modulation frequencies $\omega_M=\omega_B$, or multiple integers {$\ell$} of $\omega_B$, suggesting that tunneling occurs not only between nearest neighboring sites (i.e., when $\ell=1$) but also between sites that are $\ell\s{1}$~lattice periods apart. We assessed this phenomenon by observing an increase of the atomic distribution width following Eq.~(\ref{eq:ResTunn}).

\begin{figure}[b]
\begin{center}
\includegraphics[width=0.48\textwidth]{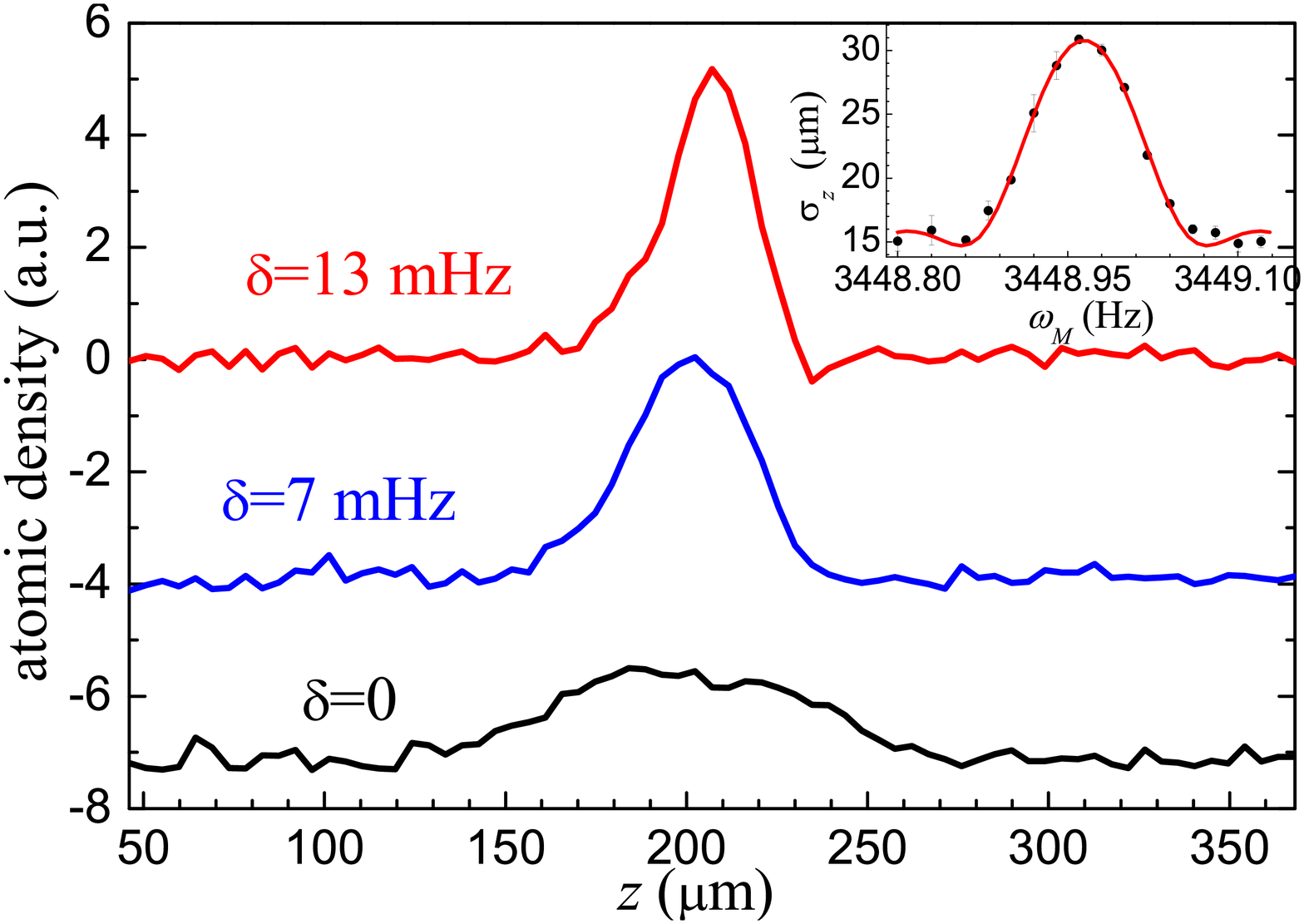}
\caption{Effect of the resonant AM technique on the atomic cloud with a modulation time of 10 s. The integrated density distributions after AM for different detunings $\delta$ from 6$\times\omega_B$ clearly appear  stretched at the resonance. In the inset, we plot the resonance spectrum derived from the rms size along the lattice $\sigma_z$ for AM resonant tunneling at the 6-th harmonic.} \label{fig:WSdeloc}
\end{center}
\end{figure}

In Fig. \ref{fig:WSdeloc} the effect of AM driving is shown: after resonant amplitude modulation at frequency $\omega_M = 6\times\omega_B$, the recorded density profile of the atomic cloud appears clearly stretched along the lattice direction, while the density height decreases.  

The measurement of the spatial rms extent along the lattice for several modulation frequencies allows to determine the Bloch frequency with high sensitivity. Long modulation times ($t_{mod}=10$ s) allow to perform intraband WS states spectroscopy with very narrow linewidths, down to Fourier-limited linewidth $\GammaF = 31$ mHz. An AM resonance spectrum for the 6-th harmonic is shown in the inset of Fig.\ref{fig:WSdeloc}. Here the observed $\Gamma$ corresponds within the experimental error to the Fourier limited linewidth $\GammaF$. Long modulation times also increase the transport resonance quality factor as the maximal extent at resonance is $v_\ell\,t_{mod}$. However, as observed in Fig.~\ref{fig:WSdeloc}, this implies a reduction of the observed optical density and thus a reduced signal-to-noise ratio of the in-situ absorption imaging of the atomic cloud. The necessary trade-off between an enhanced quality factor and a reduced signal-to-noise ratio is found by an optimal choice of the modulation depth $\alpha$. A detailed analysis of the effects which may limit the WS resonant tunneling sensitivity is presented in Sec.~\ref{sec:WSsensitivity}.

Comparable results were previously achieved with the PM resonant tunneling. Coherent delocalization was observed up to $\ell = 4$~\cite{Ivanov:2008p37} and Fourier-limited linewidths up to 15 s. In this case the existence of higher order harmonics in the lattice potential is unavoidable. The measurements show in the PM case a higher fluctuation of the broadening velocity $v_\ell$ which can be attributed to the non linear contribution of the higher order harmonics as function of the modulation strength $\beta$. For this reason and according to possible frequency shifts presented in Sec.~\ref{sec:TRS}, from now on we will consider only AM resonant tunneling technique as an accurate sensor for force measurements.

\subsection{Effects at resonant driving II: {\BOnew~experimental results}}

We experimentally compared the \BOnew~technique with the usual BO measurement for high sensitivity gravity determination. Enhancement of the interference peak during time-of-flight detection of BOs was previously used for coherent control of the spatial extent of an atomic wavefunction~\cite{Alberti:2009p45} for short times. We applied more recently this technique for high sensitivity force measurements~\cite{Poli2011}.

The phase sensitivity to BOs is greatly enhanced by means of \BOnew~described in Sec.\ref{sec:inter}. In order to demonstrate this, we employ the following sequence: we adiabatically load the atoms form the red MOT into the shallow lattice ($U_0 = 3 E_R$) so that after few milliseconds the in-lattice temperature is reduced to about 0.6 $\mu$K. We then apply an AM burst at the Bloch frequency lasting 120 cycles with a modulation depth $\alpha =0.2$. After different time intervals $t$ of free evolution in the lattice, we perform absorption imaging of the atomic cloud by means of TOF technique and the resulting momentum distribution is analyzed.

\begin{figure}[tb]
\begin{center}
  \includegraphics[width=0.5\textwidth]{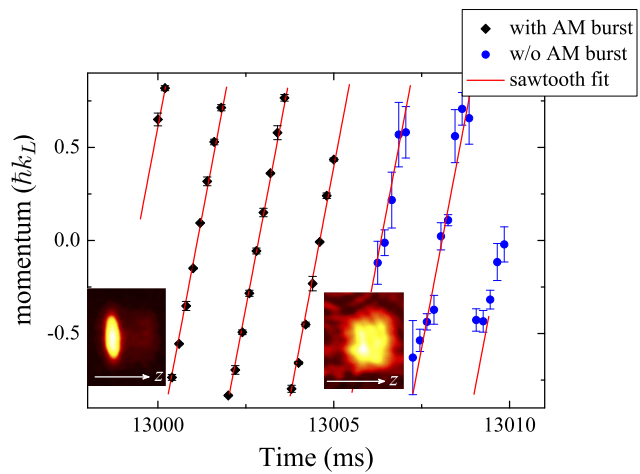}
\caption{Comparison between Bloch oscillation phase evolution with initial AM burst (black diamonds) and without it (blue circles). In the insets, 2D TOF atomic distributions are displayed for the two cases. The red line is obtained from the best fit with a sawtooth function.}
\label{fig:BOcomp}
\end{center}
\end{figure}

The result of the \BOnew~technique is shown in Fig.\ref{fig:BOcomp}: we are able to follow BOs with high sensitivity due to the formation of narrow interference momentum peak, as shown in the left inset in Fig.~\ref{fig:BOcomp}. The resulting interference pattern shows a narrow Gaussian distribution due to the convolution of the wavefunction shown in Fig.\ref{fig:Sim1} with the initial atomic spatial distribution. The effective momentum peak width is $\Delta p \approx 0.18\, \hbar k_L$ for a $t_{tof} = 14$ ms, which corresponds to $\tau_{trans}$ in (\ref{eq:dbfueirubw}). This result is nearly a factor of two narrower than the central momentum peak in non-interacting BECs \cite{PhysRevLett.100.080404,PhysRevLett.100.080405} for the same time $t_{tof}$.

The comparison with BO without initial AM burst is also shown in Fig.\ref{fig:BOcomp} (blue circles). In this case, the interference pattern is described by two overlapping Gaussian distributions overfilling the Brillouin zone~\cite{Ferrari06}. The recorded single point error $\sigma_p$ can be as large as 0.2 $\hbar k_L$, for an overall Bloch frequency relative error $\Delta\omega_B/\omega_B = 3\cdot10^{-5}$. The \BOnew~interference peak yields a typical error of each single point of the momentum evolution $\langle \sigma_p \rangle = 1.6\cdot 10^{-2}\,\hbar k_L$. This result is due to the greater visibility obtained by the AM burst technique: the peak visibility, defined as in Ref.~\cite{Gerbier:2008p18}, can be as large as 60\%, while for classical BOs it is smaller than 10\%. Implications on the precision in determining $\omega_B$ are discussed in Sec.~\ref{sec:sub_BO}.

\section{Force sensitivity and systematics of the atomic sensor}
\label{sec:PS_BO}

We analyze the sensitivity and the final precision on the determination of $\omega_B$ and its application on the determination of the gravity acceleration $g$. We proceed by summarizing the systematic effects  which affect the system for both techniques.

\subsection{Phase sensitivity in Bloch oscillations}\label{sec:sub_BO}

\begin{figure}[tb]
\begin{center}
\includegraphics[width=0.95\columnwidth]{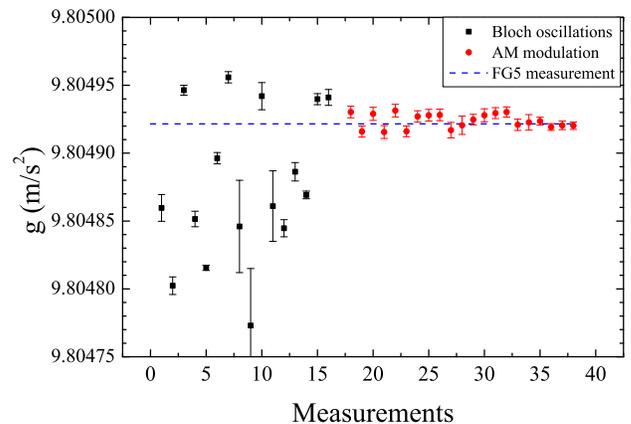}
\caption{Summary of gravity measurements by means of {\BOnew~(black squares)}~and AM {resonant tunneling (red circles)}. The dashed line represents the {precise value of the gravity $g$ measured by the FG5~\cite{Poli2011} which we use to compare the two force measurement techniques}.} 
\label{fig:BOmeas}
\end{center}
\end{figure}

Starting from an almost dispersionless momentum distribution due to the AM resonant burst, the increase of visibility yields the observation of BOs up to 17 s, mainly limited by the lifetime of the optical trap. Each gravity determination lasts more than 2 hours. Repeated measurements with the same experimental conditions as in Ref.\cite{Poli2011} are shown in Fig.\ref{fig:BOmeas}. The precision in the determination of the Bloch frequency  $\Delta\omega_B/\omega_B$ ranges between $2.2\cdot10^{-6}$ and $1.7\cdot10^{-7}$ with an observation time of about 13 s. However we observed a relative scatter of the calculated gravity values  one order of magnitude higher.  The final result of local gravity is $g_{BO}=9.80488(6)$~m/s$^2$. By looking at the residuals produced by the best fit with a  sawtooth wave, we observed a reduction of its mean value by removing the data set with longer evolution time. This is a clear signal that phase sensitivity to BO is washed out by long term drift of the apparatus. In particular, the main sources of instability which leads to this high uncertainty on the determination of gravity are the frequency of the lattice laser and the MOT center position stability, which comes from the long-term instability of the second-stage cooling laser.

\begin{figure}[b]
\begin{center}
 \includegraphics[width=0.5\textwidth]{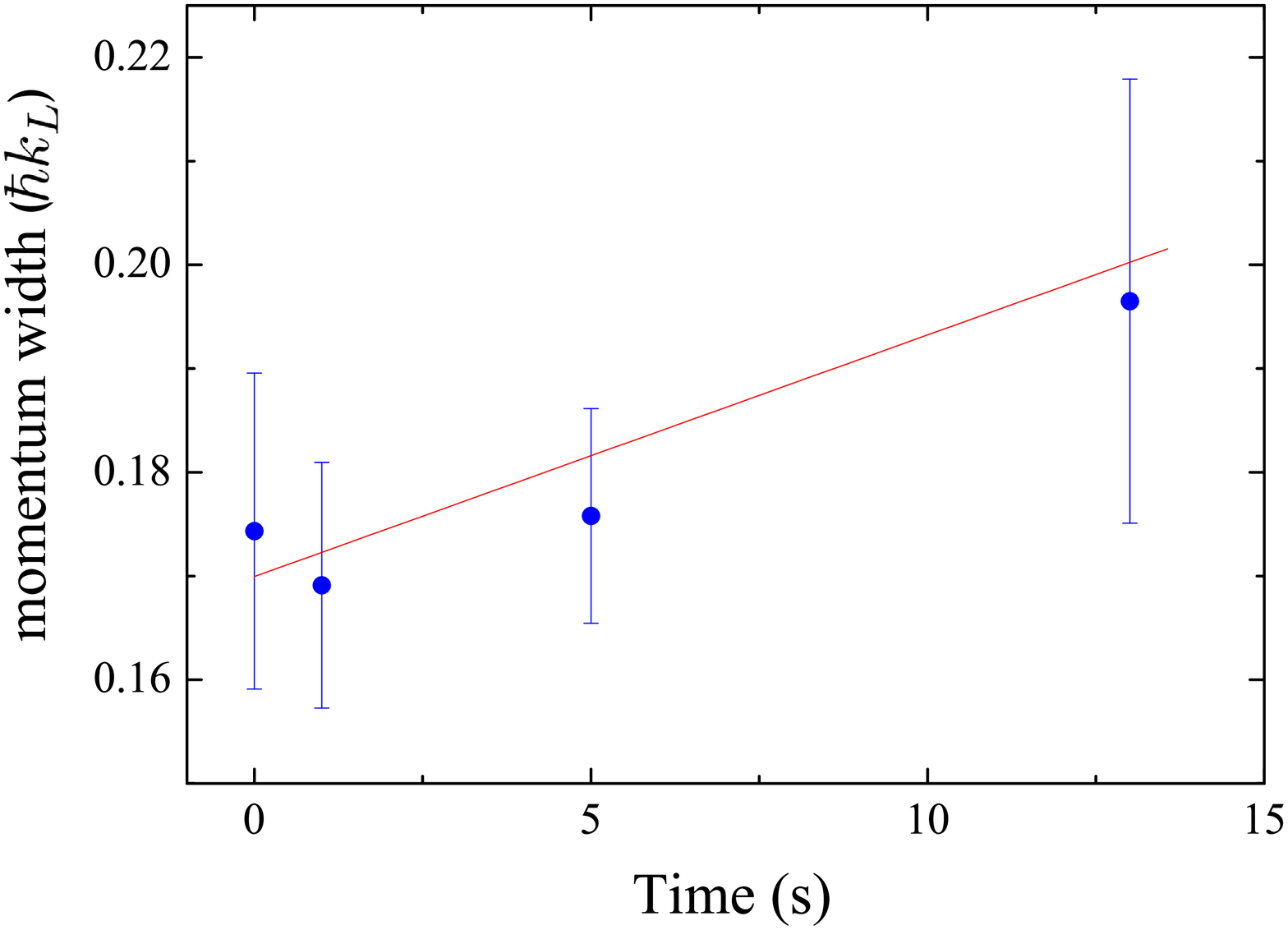}
 \caption{Broadening of the interference peak in time-of-flight measurement as function of the free evolution time inside the lattice. From the linear fit, a low density strontium cold gas can achieve a coherence time higher than 500 s.}
 \label{fig:decoh_exp}
\end{center}
\end{figure}

It is possible to determine the coherence time of the system by looking at the momentum peak broadening. According to Eq.~(\ref{eq:BO_dec2}), if the free evolution time in the lattice $t_{hold} \ll \gamma^{-1}$, we can assume a linear peak broadening $\Delta p (t)= \Delta p_0(1 +\gamma t/2) $. In fact, if we consider the decoherence due to elastic collisions, the  expected coherence time can be estimated between 440 and 880 s for a 50\% fluctuation range over the number of atom. We experimentally observed a slow increase of the interference peak width along the {13 s of measurement time}. The estimated coherence time, defined as the inverse of the slope of a linear fit, is equal to $530 \pm 150$ s, which agrees with the expected scattering rate, as resulting from the data in Fig.~\ref{fig:decoh_exp}.

A fundamental source of uncertainty on the phase of the BO measurement is the stochastic nature of the oscillation phase due to LZ tunneling during the switch-off of the lattice in a time-of-flight measurement \cite{Holthaus:2000p64}. In fact, when the atom cloud is released by the optical trap, there is a finite time interval $\Delta T_{LZ}$ in which the LZ tunneling probability is not negligible. In our experiment the switch-off time constant is about 5 $\mu$s, which roughly correspond to a non-zero LZ tunneling probability time interval $\Delta T_{LZ} =$ 14 $\mu$s and thus a random phase $\delta \phi \leq \Delta T_{LZ}\nu_B$ affecting the BO of each atom. LZ tunneling is then translated into a source of statistical uncertainty which can be numerically modeled, for instance, by means of a Poisson-like $\Delta t_{LZ}$ distribution. For our typical values of the lattice depth ($U_0 \leq 6 E_R$) and atom number ($N_{at} \simeq 5\cdot10^4$), LZ tunneling sensitivity corresponds to $\delta\omega_{LZ}/\omega_B \leq 10^{-7}$.

\subsection{Sensitivity in Wannier-Stark resonant tunneling}\label{sec:WSsensitivity}

High sensitivity measurement of resonant tunneling in an optical lattice requires coherence between the WS states to keep at a very low level any diffusive effect on the lattice which can randomly extend the atom cloud $\sigma$ thus causing a broadening of the tunneling frequency resonance $\Gamma$.

The keys to enhance the force measurement sensitivity are shown in Eq.~(\ref{eq:ResTunn}). In the case of Fourier-limited resonance \cite{Ivanov:2008p37}, the $\omega_B$ sensitivity to spatial width fluctuations $\Delta\sigma$ can be estimated from a power expansion of this formula and then by applying a first-order propagation of uncertainty theory. This results on a Bloch frequency uncertainty

\begin{equation}\label{eq:SNR_WS}
\Delta\omega_B =2\pi\times\frac{3}{\pi t^2\,v_\ell \,\ell} \ \Delta\sigma\,\, .
\end{equation}

It is noteworthy that the resonance width $\GammaF =2\pi\times(\pi t)^{-1}$ does not depend on the index $\ell$ which identifies the modulation harmonic with respect to the fundamental frequency $\omega_B$. This fact is significant because, as shown by expression~(\ref{eq:SNR_WS}) for the {uncertainty on $\omega_B$}, it allows to increase the sensitivity of the measurement by a factor $\ell$.

A typical Fourier-limited AM resonance at the 6-th harmonic has been presented in Sec.~\ref{sec:SensitivityGeneral} in Fig.~\ref{fig:WSdeloc}. In order to investigate the sensitivity of the AM resonant tunneling measurement, we performed repeated measurements of the Bloch frequency with a modulation time $t=10$ s at $\ell = 5$. Typical uncertainty of the Bloch frequency ranges between 0.1 and 0.3 mHz, which is translated into a relative uncertainty between 180 and 590 ppb.

The statistical uncertainty on the WS tunneling resonance can be estimated from (\ref{eq:SNR_WS}) by looking at the fluctuation of the atomic cloud during the cooling and trapping procedure. If one assumes an average fluctuation $\Delta\sigma/\sigma_0 \sim 2 \%$, which is a typical value for our apparatus, we get an expected relative error on the Bloch frequency of 370 ppb. This effect represents the main source of statistical uncertainty for the WS resonant tunneling. It can be reduced by starting atomic cloud fluctuations by means of a narrow optical tweezer~\cite{SorrentinoPRA}.

Any source of broadening of the resonance is translated into a smaller quality factor and thus into a reduced sensitivity to the external force sensed by the atoms. In resonant tunneling experiments, collisions lead to sequential tunneling instead of coherent {tunneling}. Sequential tunneling is expected to increase the width of resonance spectra with respect to the Fourier limit.

\begin{figure}[tb]
\begin{center}
 \includegraphics[width=0.45\textwidth]{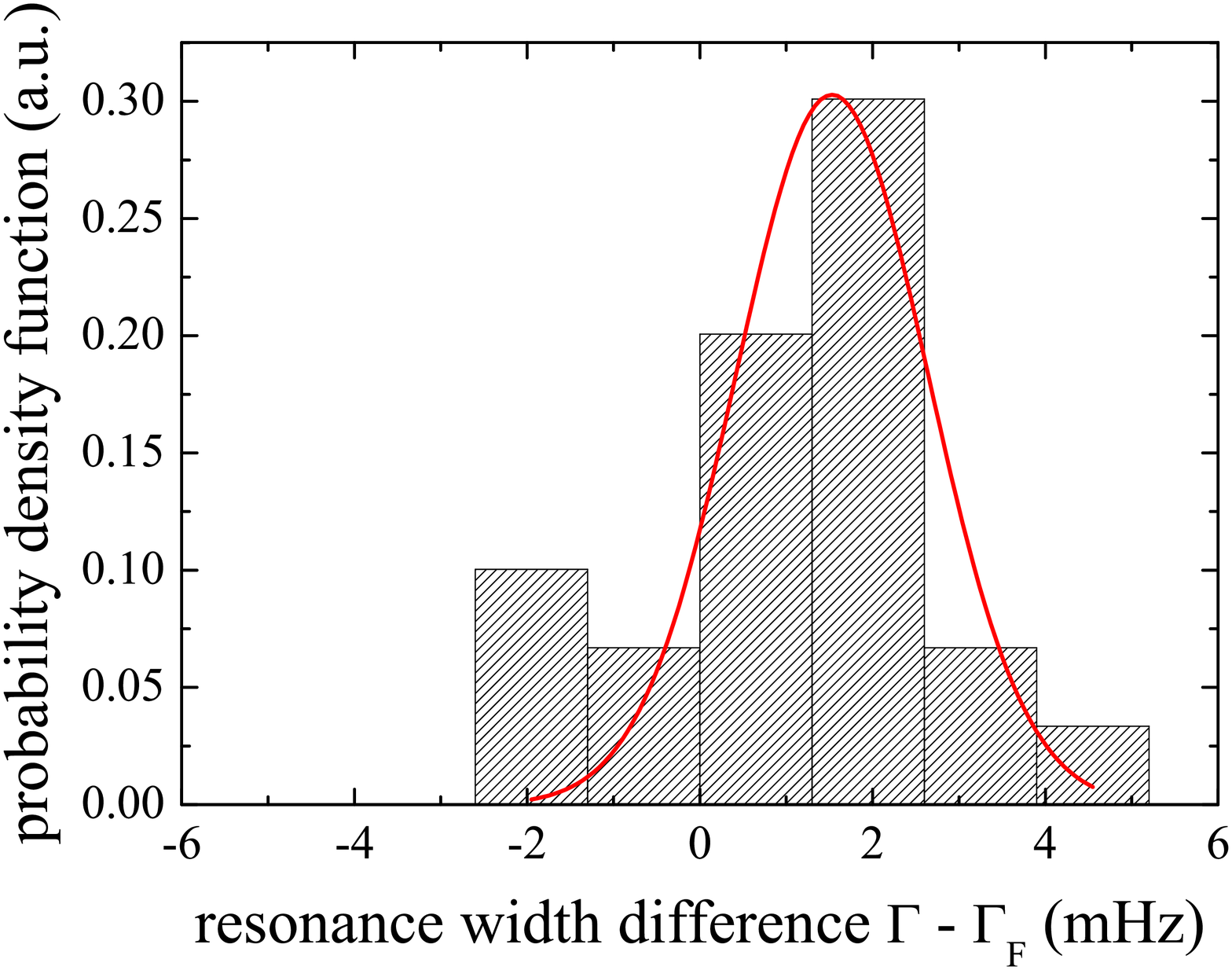}
 \caption{Study of the broadening of the AM tunneling resonance. {Considering} a modulation time of 10 s, {which corresponds to a} relative Fourier linewidth {of} 31 mHz, it is possible to estimate the coherence time of the system {by measuring the deviation of the experimental linewidth from its fundamental limit}.}
 \label{fig:decoh_exp2}
\end{center}
\end{figure}

We analyzed the values of $\Gamma$ that we extracted from each fit used for Bloch frequency determination. A histogram of the deviations from the Fourier-limited resonance $\GammaF$ is reported in Fig.~\ref{fig:decoh_exp2}. Here one can observe a Gaussian-like distributed broadening of the AM resonance, with a mean value of about 1.5 mHz above the Fourier limit and a standard deviation of 1.3 mHz. If one considers a maximum line broadening of 2.8 mHz and assumes this estimate as the decoherence rate, it means that the coherence time is better than 330 s, which is again consistent with collisional effect and with the BO's estimate as represented in Fig.\ref{fig:decoh_exp}.

Another source of sequential tunneling can be due to spurious AM . As shown in Sec.~\ref{sec:ExperimentalSetup}, the servo-loop system controlling the lattice laser intensity has been optimized in order to avoid spurious amplitude modulation . Residual amplitude noise on the lattice laser intensity induces resonant tunneling at the harmonics of the Bloch frequency. The level of the residual relative intensity noise is about -120 dB, which yields a fluctuation on the atomic size $\Delta\sigma_{RIN} \simeq 10^{-2} \mu$m. This effects may limit the sensitivity on the Bloch frequency at the level of 2$\times$10$^{-8}$ in relative units.

\subsection{Use as a force sensor: systematic effects}

We analyze in this section the systematic effects which affect the measurement of the gravity acceleration $g$. These effects are common to both BO measurement and WS resonant tunneling techniques.

\subsubsection{Lattice light shifts}

\begin{figure*}[bt]
\begin{center}
\includegraphics[width=0.95\textwidth]{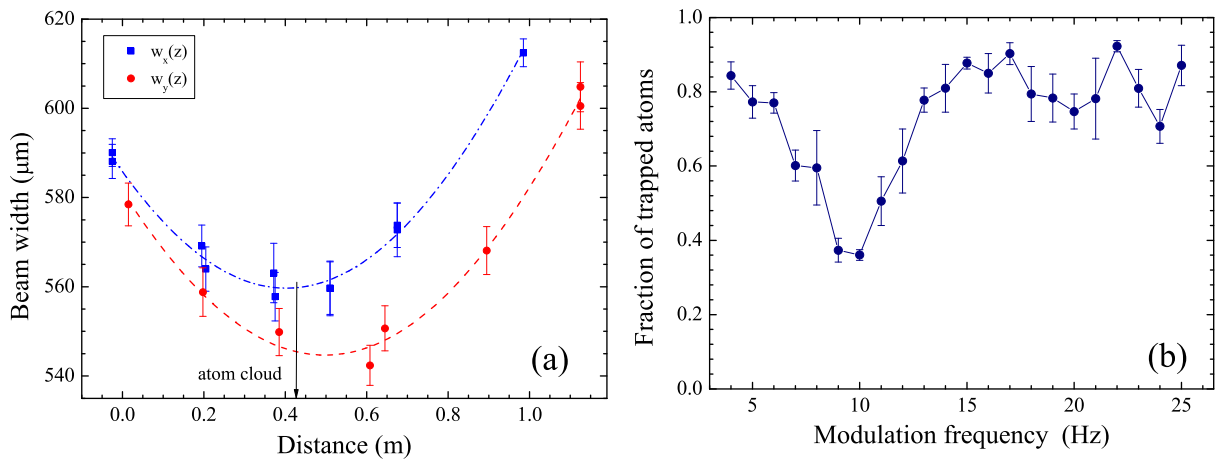}
\caption{Characterization of the lattice beam propagation and intensity in the proximity of the atomic cloud. (a) The measurement of the beam width of two orthogonal transverse directions (respectively squares and circles) determines the trapping potential gradient (best fits are respectively the (blue) dot-dashed and the (red) dashed curves), while (b) parametric excitation of radial motion is used to calibrate the intensity of the optical lattice at the place of the atoms.} \label{fig:calibration}
\end{center}
\end{figure*}

The main contribution to systematic shift in local gravity measurement with trapped neutral atoms is due to the lattice light itself.  The spatial inhomogeneity of both the intensity and the wave vector of the Gaussian lattice beam yields space-dependent variations of the potential energy, which results in

\begin{equation}\label{eq:Utotnew}
U_{tot}(z) = U_{s}(z)+U_{l}(z)\cos(2k(z)z)-mgz\ ,
\end{equation}
where $U_s$ and $U_l$ depend respectively on the squared sum and on the product of the electric field amplitudes of the two counter-propagating beams. Comparing this external potential with the ideal lattice in (\ref{eq:Uz}), besides the gravitational potential one recognizes another term due to the lattice intensity gradient. In the case of a red detuned lattice laser, atoms' wavefunctions are mainly located at the antinodes of the lattice where the intensity (and the potential) is maximum,  so that the driving force is $F=-\partial_z V(z)=\partial_z(U_{s}(z)+U_{l}(z)-mgz)$. Furthermore, a change in the lattice constant $d = \pi/k(z)$ must be considered due to the Gouy effect \cite{Clade09}, that is
\begin{equation}\label{eq:Gouy}
k(z) = k_L-\frac{d}{2dz}(\Phi_r(z)-\Phi_i(z))\bigg|_{z=z_0}
\end{equation}
where $\Phi_i(z)$ and $\Phi_r(z)$ are the phases for the incoming and reflected beams, respectively, and $z_0$ is the position of the atomic cloud. Thus, the gravity value becomes

\begin{equation}\label{eq:STARK}
g=\frac{1}{m}\left(2\hbar k_L\omega_B-\frac{\partial U_{OL}(z)}{\partial
  z}-\hbar\nu_B
\frac{d}{dz}\Delta\Phi_{Gouy}(z)\right)\bigg\vert_{z=z_0}
\end{equation}

\noindent where $U_{OL}(z) = U_s(z) + U_l(z)$ is lattice field depth and $\Delta\Phi_{Gouy}(z)$ is the difference of the Gouy phase of the two beams. The Gaussian beam nature of the lattice introduces two extra terms in (\ref{eq:STARK}) due to the intensity gradient and the Gouy phase shift that we call respectively $\Delta g_U$ and $\Delta g_k$. These sources of systematic shift must be estimated for a precise determination of the local gravity value $g$. We accomplished this task by a precise determination of both the geometry of the incoming and reflected trapping beams and the position of the cloud with respect beam waist, with a relative uncertainty of 1\%, as show in Fig.~\ref{fig:calibration}(a). From this measurement we estimated a beam size on the atomic cloud $w(z_0) = 557(7)~\mu$m, considering also the beam ellipticity due to different beam sizes in the two transverse directions.

An independent determination of the transverse beam size at the atomic sample position has been performed by measuring the radial atomic trap oscillation frequencies through parametric heating technique~\cite{Savard97,Jauregui01}. On Fig. \ref{fig:calibration}(b) we recorded the radial characteristic frequency $\nu_R = 5.3(1)$ Hz for an input laser power $P_{in}$ = 0.848 W. Since the relation between these parameters is known \cite{Jauregui01}, we can express the potential depth $U_0$ as function of the input power as

\begin{equation}\label{eq:LattShift}
U_0=k P_{in},\qquad \mbox{and} \qquad U(z)= U_0 I(z)
\end{equation}
where $I(z)$ is the normalized intensity profile given by $w$ and $z_0$.  For typical experimental parameters the two terms are $\Delta g_U  = 1.53(3)\cdot10^{-5}$m/s$^2$ and $\Delta g_k = 1.0(2)\cdot10^{-8}$m/s$^2$.

It is possible to reduce systematic effects from the intensity gradient by employing a blue-detuned lattice laser~\cite{Clade09}. In this case, atoms are located at the intensity nodes where the light intensity is minimal so that the second term in (\ref{eq:Utotnew}) can be neglected. Since $U_s \simeq U_l$ typically, the blue-detuned lattice light shift is  reduced by a factor of two.

\subsubsection{Other sources of systematics and final uncertainty budget}

Beside the light shifts and the calibration of the lattice frequency, there are other technical and {fundamental} sources of shift of the Bloch frequency.

As shown in the experimental apparatus description, during gravity measurement the vertical alignment of the lattice beam has been checked for each measurement.  Moreover, during the measurement a tilt-meter with a resolution of 1.7~$\mu$rad and attached to the optical table has been employed to check alignment stability.  This procedure has typical errors of 0.5 mrad leading to a relative uncertainty $\delta g/g=10^{-8}$ while the stability of the optical table over the duration of a typical measurement ($\leq 1.5$ hour) is better than 10~$\mu$rad.

We estimated the contribution due to the distribution of masses around the experimental setup and we found that the biggest contribution to the local field is given by the 3.5 $\times$ 1.5 m optical table (20 cm thickness) that supports the experiment.  Given the distance between the table surface of 25 cm and the weight of 640 kg we calculated the effect on the vertical direction to be $\Delta g=2\cdot10^{-8}$ m s$^{-2}$ or ($2\cdot 10^{-9}$ in relative units). This effect is far below the current sensitivity.

\begin{table*}[tb] 
   \centering 
   \begin{tabular}{l@{\hspace{0.5cm}}c@{\hspace{0.5cm}}c}
   \toprule\\[-3mm]
     Effect   & {Correction} (10$^{-7}$) & Uncertainty (10$^{-7}$)\\[0.5mm]
     \toprule\\[-3mm]
     Lattice wavelength fluctuations     & 0 & 2 \\
     Lattice beam vertical alignment & 0 &  0.1 \\
     Inhom.\ Stark shift (beam geometry)&     14.3 $\div$ 17.3 & 0.4 \\
     Experiment timing  & 0 & 0.2\\
     Tides & -1.4 $\div$ 0.9 & $<$0.1\\
     Off-resonance tunneling   & $<$0.01 &  0.2 \\[0.5mm]
      \hline\\[-3mm]
     Systematics total & 17.2 $\div$ 22.5  &  2.1\\
     (w\s{-0.8}/\s{-0.8}o lattice wavelength fluct.) & \rule[0.9mm]{0.22mm}{1mm}\hspace{1pt}\rule[0.9mm]{0.22mm}{1mm} & 0.50\\[0.5mm]
      \toprule
    \end{tabular}
    \caption{Uncertainty budget concerning the gravity measurement with the atomic gravimeter. Correction values and their uncertainties depend on operating conditions. Typical values are given.}
    \label{syst}
  \end{table*}

With the increase of the harmonic index {$\ell$}, it is possible that for WS resonant tunneling measurements, at high modulation depths the contribution of off-resonant transitions has to be taken into account. The overlap of an infinite series of WS transitions induces a resonance asymmetry into the observed frequency spectrum and a consequent shift of the resonance peak. However the shift scales as $(\omega_B^2\,t^{3})^{-1}$ and it is fully suppressed below $10^{-10}$ level. However, this effect couples with residual amplitude noise, which we already estimated to limit the sensitivity at the level of  20 ppb. Furthermore, off-resonant interband transitions can be non-negligible \cite{Gluck:2000p318} and with expected linewidths of the order of kHz. We experimentally checked possible shifts and broadening of the resonance due to off resonant couplings and higher order effects by changing the modulation depth from 4\% to 10\%. Within the statistical error in the determination of the resonance center and resonance linewidth we do not observe any shift or deviation from the Fourier limit in the measured linewidth.

Another effect to be taken into account for these measurements comes from tidal forces. The peak-to-peak effect of tides at our site is of the order of $2 \times 10^{-6} \mathrm{m/s^2}$, i.e., $\Delta\omega_B/\omega \leq 100$ ppb. This value is only two times smaller than the current best sensitivity for $\omega_B$ measurement. Nevertheless the correction for this effect must be applied in accurate gravity measurements \cite{Poli2011}. Since each measurement lasts about 1 hour, the variation of $g$ during every measurement due to tides is below $10^{-7} \mathrm{m/s^2}$ (i.e., below 10 ppb in relative units), then Earth tides does not affect the current uncertainty budget.


In table \ref{syst} we present a summary of all the important systematic effects occurring in a gravity measurement are summarized. Any Bloch frequency determination is currently limited by 200 ppb uncertainty due to systematic effects, mainly dominated by the lattice frequency uncertainty. Beside this effect, the force sensor based on WS resonant tunneling has proven a systematic error at level of 50 ppb. According to (\ref{eq:SNR_WS}), if a frequency stabilized lattice laser is employed, this level of sensitivity can be achieved after 26 s of modulation with the current setup at the 6-th harmonic.  

\section{Conclusions}\label{sec:Conclusions}

We explored the physical properties and the technical problems behind the measurement of forces by atoms trapped in driven optical lattices. After fifteen years from the first detection~\cite{BenDahan:1996p31} of the quantum phenomenon of BO's, the precision and the accuracy of force measurement have reached the hundreds of ppb-level. This has been possible by exploiting the delocalization obtained by dynamical mechanisms and the matter-wave interference nature of this phenomenon. In particular, we have shown that the \BOnew~measurement originates from modulation-induced delocalization leading to dispersionless interference peak in momentum space~\cite{Alberti:2009p45}.

Direct measurement of the Bloch frequency, and thus of the external forces, can be performed by means of \emph{in-situ} measurements of the atomic cloud dynamics under AM or PM resonant tunneling among WS states. In particular, our experimental and theoretical analysis shows that the AM resonant tunneling results the most accurate technique with higher potential for further improvement both in accuracy and in sensitivity. We provide a detailed characterization of the uncertainty budget of our force sensor. In the base setup, its uncertainty is due to the lattice wavelength fluctuations. A frequency stabilized lattice laser would allow accurate measurements of forces with 50 ppb uncertainty. This value approaches performance demonstrated in free-fall atom gravimetry~\cite{Peters99} with the advantage of a small interferometer suited to microscale experiments.

\begin{acknowledgments}
{This research is carried out within the project iSense, which acknowledges the financial support of the Future and Emerging Technologies (FET) programme within the Seventh Framework Programme for Research of the European Commission, under FET-Open grant number 250072. This work is also supported by INFN and LENS (Contract No. RII3 CT 2003 506350). A.~A. acknowledges support from the Alexander von Humboldt Stiftung. We thank D.\ Sutyrin for his help on lattice frequency measurements. We thank G. Ferrari for a critical reading of the manuscript.}
\end{acknowledgments}

\appendix

\section{Phase-modulation Hamiltonian}\label{appendix:phasemod}
In the rotating frame the Hamiltonian in Eq.~(\ref{eq:Hpm_start}) becomes
\begin{equation}
	\C{H}_{\mathrm{PM}}'=\C{U}_\textrm{RF}\hspace{1pt}\C{H}_{\mathrm{PM}}\hspace{1pt}\C{U}_\textrm{RF}^\dag-i\hbar\hspace{3pt}\C{U}_\textrm{RF}\hspace{-1pt}\frac{\mathrm{d}}{\mathrm{d}t}\C{U}_\textrm{RF}^\dag\,.
\end{equation}
The resulting expression
\begin{multline}\label{eq:jufbdfjfbuivf}\raisetag{26.3mm}	\C{H}_{\mathrm{PM}}'=\hspace{-7pt}\sum_{n=-\infty}^{+\infty}\hspace{-4pt}n\frac{\hbar\hspace{0.5pt}\delta}{\ell} \spc\Psi_n\ket\bra\Psi_n\spc\hspace{2pt}-\hspace{-5pt}\sum_{n=-\infty}^{+\infty}\hspace{-5pt}
\Big\{\beta\hspace{1pt}d\hspace{2pt}\C{C}_\ell^{\mathrm{PM}}\s{1}\times\\
\times\sin(\omega_Mt\s{-1}-\s{-1}\phi) \exp\s{-1}\big[i \hspace{0.6pt}\beta\hspace{0.3pt}\ell\hspace{0.3pt}d/(\hbar\hspace{0.3pt}\omega_M)\cos(\omega_Mt-\phi)\s{-1}-\s{-1} i\hspace{0.6pt}\omega_Mt\big]\s{1}\times\\[1.5mm]
\times\spc \Psi_{n+\ell}\ket\bra \Psi_{n}\spc+\mathrm{h.c.}\Big\}
\end{multline}
can be further simplified by making use of the Jacobi-Anger expansion and by keeping then only the time-independent resonant terms
\begin{multline}\label{eq:jufbddfhdbfjfbuivf}	\C{H}_{\mathrm{PM}}'=\hspace{-6pt}\sum_{n=-\infty}^{+\infty}\hspace{-4pt}n\frac{\hbar\hspace{0.5pt}\delta}{\ell} \spc\Psi_n\ket\bra\Psi_n\spc+
\hspace{-4pt}\sum_{n=-\infty}^{+\infty}\hspace{-4pt}
\bigg\{\hspace{-1pt}\beta\hspace{1pt}d\hspace{2pt}\frac{\C{C}_\ell^{\mathrm{PM}}}{2}e^{i\hspace{0.5pt}(\pi/2-\phi)}\,\times\\
\times \hspace{1pt}
\left[J_0\left(\frac{\beta\hspace{0.3pt}d\hspace{1pt}\ell}{\hbar\hspace{0.5pt}\omega_M}\right)+J_2\left(\frac{\beta\hspace{0.3pt}d\hspace{1pt}\ell}{\hbar\hspace{0.5pt}\omega_M}\right)\right]
\spc \Psi_{n+\ell}\ket\bra \Psi_{n}\spc+\mathrm{h.c.}\hspace{-1pt}\bigg\}\,.
\end{multline}
In order to retrieve the Hamiltonian in Eq.~(\ref{eq:Hpm}) one has to use the fact that $J_0(x)+J_2(x)=(2/x)J_1(x)$ and that
\begin{equation} \label{eq:tunn_rates_PM} 
\bra \Psi_{n+\ell}\spc z\spc \Psi_{n}\ket/d=\left\{\s{-1}
\begin{array}{lcl}
	n&\hspace{0.3mm}\textrm{for}&\hspace{0.6mm}\ell=0\\[1mm]
	{\varepsilon_\ell}/{\hbar\hspace{0.4pt}\omega_B}&\hspace{0.3mm}\textrm{for}&\hspace{0.6mm}\ell\neq0
\end{array}\right.
\end{equation}
with the coefficients $\varepsilon_\ell$ defined in Eq.~(\ref{eq:epsilon_coeff}).

\section{Time-evolution operator}\label{appendix:timeevoloper}
It is convenient to express the WS states $\spc \Psi_n\ket$ in terms of the states defined in  Eq.\ (\ref{eq:WSdelta}), which are eigenstates of the Hamiltonians in Eqs.\ (\ref{eq:WS_AMham}) and (\ref{eq:Hpm}):
\begin{equation}
	\spc\Psi_n\ket= \hspace{-6pt}\sum_{m=-\infty}^{\infty}\hspace{-5pt}e^{-i\s{0.6}m\s{0.7}(\pi/2-\phi)}\s{1}J_{-m}\s{-2}\left(\s{-1}\frac{\C{J}_\ell}{F_\delta\s{0.7}d}\s{-1}\right) \spc \Phi_{n-m\s{0.5}\ell}\s{0.3}(\delta)\ket
\end{equation}
The time-evolved state at time $t$ is given by
\begin{multline}\raisetag{20mm}
	\spc \Psi_n(t)\ket=\s{-8}
	\sum_{m=-\infty}^\infty\s{-5}\bra\Phi_m(\delta)\spc\Psi_n\ket\s{1.5}e^{-i\s{0.3}m\s{0.5}\delta\s{0.4}t/\ell}\s{1.5}\spc\Phi_m(\delta)\ket=\\
	=\s{-1}e^{-i\s{0.3}n\s{0.3}\delta\s{0.3} t/\ell}\s{-8}\sum_{p,q=-\infty}^\infty\s{-8}
	e^{-i\s{0.3}p\s{0.3}(\pi/2+\phi)}
	e^{i\s{0.3}q\s{0.3}\delta \s{0.3}t}
	J_{q}(x)\s{1}
	J_{q+p}(x)
	\s{1}\spc\Psi_{n+p\ell}\ket\s{-1}
\end{multline}
with $x={\C{J}_\ell}/({F_\delta\s{0.7}d})$.. By using the addition theorem of Bessel function, which states
\begin{equation}
\label{eq:dnfudfeoesodbdddfdkdfjkf}
\sum_{q=-\infty}^{\infty}\hspace{-3pt}e^{i\s{0.3}q\s{0.3}\alpha}J_q(x)\s{0.5}J_{q+p}(x)=e^{i\s{0.3}p\s{0.3}(\pi-\alpha)/2}\s{1}J_p\s{-0.5}\big[2\hspace{1pt}\sin(\alpha/2)\,x\big],
\end{equation}
one finds directly the expression in Eq.\ (\ref{eq:WS_time}).

\section{Ballistic expansion at resonance}\label{appendix:sigma}

Considering the special case on resonance, $\delta=0$, the size $\sigma(t)$ of an atomic wave packet can be calculated in the asymptotic limit by means of the group velocities $v_g(k)$, which are defined, analogously to Eq.\ (\ref{eq:vg}), as
\begin{equation}
v_g (k)=\frac{1}{\hbar}\frac{\partial E_{\mathrm{mod}}(k)}{\partial k}=\frac{\ell\s{0.5}d\s{0.5}\C{J}_\ell}{\hbar}\cos(\ell\hspace{0.5pt} k\hspace{0.5pt}d+\phi)\,,
\end{equation}
where $E_\mathrm{mod}(k)$ is the energy dispersion relation derived in Eq.~(\ref{eq:energy}) in the presence of resonant driving. Taking into account the uniform momentum distribution of the initial state $\spc\Psi_n\ket$, i.e., $\spc\bra k\spc\Psi_n\ket\spc^2=1$, the variance at time $t$ results
\begin{equation}
	\sigma_{\Psi_n}^2(t)\approx\s{-3}\int_{-\infty}^{\infty}\s{-7}\mathrm{d}x\s{3}x^2\left(\frac{d}{2\pi}\int_{\BZ}\s{-2}\mathrm{d}k\s{3}\delta[x-v_g(k)\s{0.3}t\s{0.3}]\right)= v_{\ell}^2\s{0.8}t^2
\end{equation}
in agreement with Eq.\ (\ref{eq:ResTunnAA}). Since the computation of $\sigma_{\Psi_n}^2(t)$ is rather lengthy, we shall report only the final result for the AM modulation,

\begin{multline}
\sigma_{\Psi_n}^2(t) = v_{\ell}^2\s{0.8}t^2\,\sinc\s{-1}  \left[\delta/\GammaF(t)\right]^2+2\sqrt{2}\s{2}v_\ell\s{0.3}t\s{1.5}d\s{2.5}\epsilon_\ell/(\hbar\omega_B)\times \\[1mm]\times\cos\left[\phi-(\omega_M+\ell\omega_B)\s{0.5}t/2\right]\sinc\left[\delta/\GammaF(t)\right]+\sigma^2_{\scalebox{0.8}{$\Psi_n$}}(0),
\end{multline}
with the coefficients $\varepsilon_\ell$ defined in Eq.~(\ref{eq:epsilon_coeff}), $\GammaF(t)$ the resonance linewidth, and $\sigma_{\scalebox{0.8}{$\Psi_n$}}(0)$ the size of the initial state $\spc\Psi_n\ket$.

\bibliographystyle{apsrev}
\bibliography{biblio_final.bib}

\end{document}